# Frequency-dependent Phonon-mediated Unidirectional Magnetoresistance in a Metal on an Insulator with Highly Nonequilibrium Magnons

Sean E. Sullivan,[1,+,^] Hwijong Lee,[1,^] Annie Weathers,[2,#] Li Shi[1,2,*]

[1]Materials Science and Engineering Program, The University of Texas at Austin, Austin, Texas 78712

[2]Department of Mechanical Engineering, The University of Texas at Austin, Austin, Texas 78712

+Current address: Materials Science Division and Center for Molecular Engineering, Argonne National Laboratory, Lemont, Illinois 60439, USA

#Current address: MIT Lincoln Laboratory, Lexington, Massachusetts 02421, USA

^These authors contribute equally

*Author to whom correspondence should be addressed: lishi@mail.utexas.edu

Heavy metal (HM)/magnet bilayers host many magnetoresistances (MR) and spin caloritronic effects. Here we show that the spin Peltier effect and electron-phonon scattering produce much larger unidirectional MR of an HM on a magnetic insulator than existing theories that neglect the interplay between MR and spin caloritronic effects. By accounting for local nonequilibrium in both the magnon chemical potential and temperature, our analytical model attributes the observed frequency dependence of the spin Peltier MR and the spin Seebeck effect to the reduction of the thermal penetration depth, which approaches the 1-μm scale magnon spin diffusion length at high frequencies.

A panoply of magnetoresistance (MR) [1–3] and spin caloritronic effects [4–7] has emerged in heterostructures comprising a heavy metal (HM) and a ferromagnetic metal or insulator (FM or FI). These effects enable the conversion among electrical, thermal, and magnetic signals for information processing and energy conversion. Compared to the well understood mechanisms for anisotropic magnetoresistance [8] and giant magnetoresistance (GMR) [9,10], the physics behind the recently observed MR effects in HM/FM and HM/FI heterostructures is nuanced. A previous experiment [1] observed a minimum in the resistance of an HM film on an FI when the in-plane magnetization (*M*) of the FI was made perpendicular to the electric current (*I*) in the metal. Dubbed as a spin Hall magnetoresistance (SMR), this minimum has been attributed to the interplay between the spin Hall effect (SHE) and the inverse SHE (ISHE) [11–14], where the spin current generated by the SHE is reflected by the interface to produce an additional *I* component via the ISHE when the net spin polarization ($\sigma$) in the HM is collinear with *M* [1]. HM/FM bilayers further exhibit a magnetoresistance that changes sign upon *M* reversal [2]. Referred to as a unidirectional MR (UMR), this breaking of time reversal symmetry has been explained based on spin-dependent electron-electron scattering analogous to current-in-plane GMR, except that the role of the FM polarizer in GMR is replaced by the SHE in the normal metal [2,15,16]. Meanwhile, a UMR has also been attributed to the variation in the interface spin flux caused by a change in the FM magnon accumulation upon magnetization switching [15,17]. By examining the UMR in Pt on yttrium iron garnet (YIG), a theoretical study investigated the electron-magnon scattering mechanism and predicted a small UMR on the order of $10^{-8}$ in this prototypical HM/FI heterostructure [18].

In parallel, the spin Seebeck effect (SSE) has been explained based on the ISHE-mediated conversion of a thermally driven interfacial spin flux into an in-plane electric field in a metal on an FI [4]. Conversely, temperature modulation from the spin Peltier effect (SPE) is attributed to the heat transported by an interfacial spin current injected by the SHE from an in-plane electric current [5]. Often neglected in the studies of various MR effects in the same heterostructures, the SPE and SSE contributions are controlled by coupled spin and heat transport properties of nonequilibrium magnons [5,19], measurements of which are challenging due to the coexistence of large phonon contributions to heat transport.

Here we report measurements of a UMR on the order of $10^{-6}$ in a Pt/YIG heterostructure, two orders of magnitude larger than previously predicted [18]. The result is described as a spin Peltier MR (SPMR) due to cooling and heating of the phonons in the HM by the spin Peltier effect [5,6,20] together with electron-phonon coupling, both of which have not been accounted for in prior UMR theories. We consider both the magnon chemical potential and magnon-phonon temperature nonequilibrium in an analytical solution of the coupled spin, charge, and heat transport equations. Our analysis is able to explain the observed frequency dependence of the SPMR and the reciprocal SSE based on the interplay among the frequency-dependent thermal penetration depth ($\Lambda$), a 1-µm scale magnon spin diffusion length ($\lambda$), and a magnon-phonon thermalization length ($l_{mp}$) that is previously reported to be below 300 nm [21–24].

Our samples comprise 5.8-nm-thick Pt Hall bar structures patterned on 3-µm-thick YIG films on a $Gd_3Ga_5O_{12}$ (GGG) substrate, which was kept at room temperature during the measurements. When a sinusoidal current with root-mean-square (rms) amplitude $I_{\omega_J}$ was injected at a frequency $\omega_J/2\pi$, the longitudinal resistance of the Pt line measured with a lock-in amplifier at the first harmonic ($1\omega_J$) frequency exhibited the lowest value when $\boldsymbol{M}$ was collinear with the current, see Fig. S2 in the Supplementary Materials [25]. Both the $\boldsymbol{M}$ sign dependence and the $10^{-4}$ magnitude of the observed MR are consistent with prior SMR measurements [1], yielding an interfacial spin mixing conductance ($g_r$) of $1.1 \times 10^{14}\ \Omega^{-1}\mathrm{m}^{-2}$.

In addition to the SMR measured at $1\omega_J$, we detected a second harmonic ($2\omega_J$) out-of-phase transverse voltage signal, $V^T_{2\omega_J,Y}$, that switched sign together with $\boldsymbol{M}$ when $\boldsymbol{M}$ is collinear with the current at $\varphi = 180°$, as shown in Fig. 1(a, b). This signal resembles the SSE reported for this geometry [26]. When the in-plane $\boldsymbol{M}$ was rotated to $\varphi = -90°$, $V^T_{2\omega_J,Y}$ vanishes in Fig. 1 (c), in agreement with the expected orthogonality between $\boldsymbol{M}$ and the electric field due to the SSE. Here,

X and Y are the in-phase and out-of-phase components of the lock-in amplifier-detected signal, as compared to a sinusoidal reference at the corresponding harmonic.

In contrast, the second harmonic longitudinal voltage signal ($V^L_{2\omega_J,Y}$) switched sign with $M$ at $\varphi = -90°$ [Fig. 1(d, f)] and vanished for $\varphi = 180°$ [Fig. 1(e)]. In the longitudinal measurement, we utilized a balanced circuit to nullify the large $1\omega_J$ voltage component. Taking advantage of the device symmetry to reduce common mode noise, we injected the sinusoidal current through the central contacts and measured the longitudinal voltage as current flowed to both the left and right sides of the Pt line.

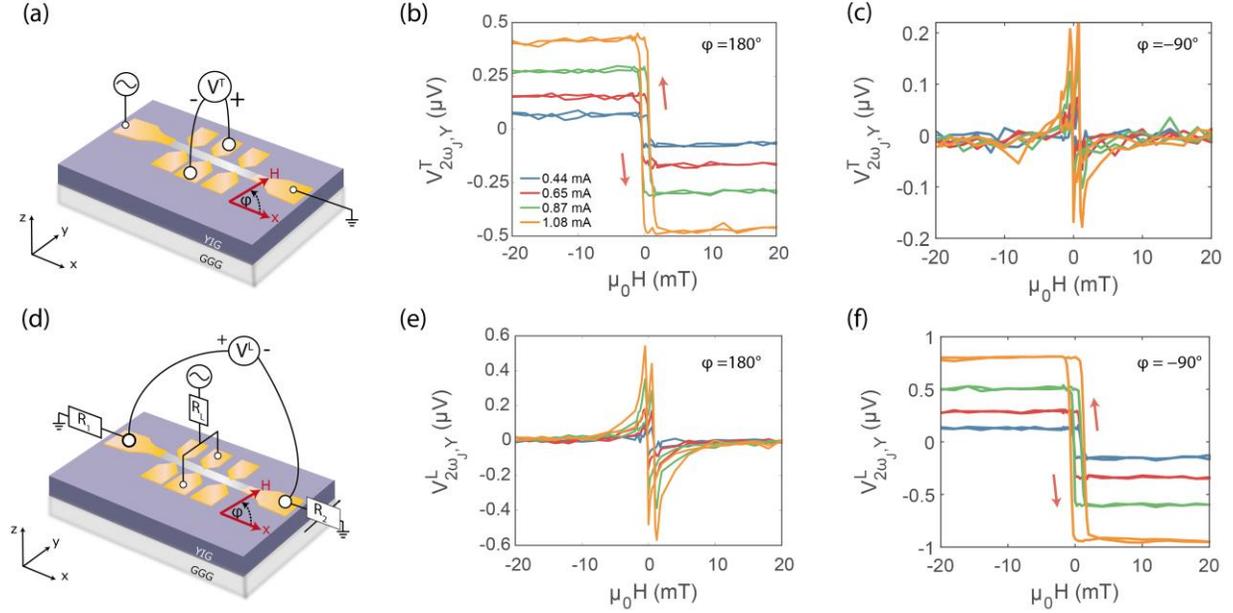

**FIG. 1.** Second harmonic transverse and longitudinal electrical measurements. (a) Schematic illustration of the transverse lock-in electrical measurement configuration. (b) Second harmonic out-of-phase voltage component, $V_{2\omega_J,Y}$, measured by the middle transverse Hall probes as a function of the applied magnetic field (H) that is collinear with the direction of the electric current, $\varphi=180°$. (c) Transverse $V_{2\omega_J,Y}$ for $\varphi=-90°$. (d) Longitudinal electrical measurement configuration, where a precision resistor ($R_1$) and variable resistor ($R_2$) are used in a balanced circuit to nullify the $1\omega_J$ component of the bridge voltage $V^L$. $R_L$ is the output impedance (50 Ω) of the voltage source of an ultrahigh frequency lock-in amplifier. (e) The longitudinal $V_{2\omega_J,Y}$ for $\varphi = 180°$ and (f) longitudinal $V_{2\omega_J,Y}$ for $\varphi = -90°$. The root-mean-square (rms) amplitude of the sinusoidal electric current through the Pt line increases from 0.44 mA (blue) to 1.08 mA (purple)

supplied at a frequency $\omega_J/2\pi = 50.17$ Hz. The H field-independent background has been subtracted from all $V_{2\omega_J,Y}$ data.

Figure 2(a) shows the changes, $\Delta V^L_{2\omega_J,Y}$ and $\Delta V^T_{2\omega_J,Y}$, in the second harmonic voltage as the magnetic field was swept from $-20$ to $+20$ mT for both the longitudinal configuration at $\varphi = -90°$ and the transverse configuration at $\varphi = 180°$, respectively. The magnitudes of these Y components [Fig. 2(b)] decrease while the X components [Fig. 2(c)] increase with increasing $\omega_J$. Bridge circuit analysis produces an SSE contribution to the measured $\Delta V^L_{2\omega_J,Y}$ as $-\gamma_L(L_h/w_H)\Delta V^T_{2\omega,Y}$, where $L_h = 320$ μm is the length of each half of the Pt line, $w_H = 29.5 \pm 1.5$ μm is the measured distance between the transverse Hall probes, and $\gamma_L$ depends on the resistances in the bridge circuit (see Supplementary Materials section H [25]). The apparent difference between the measured $\Delta V^L_{2\omega_J,Y}$ and the obtained SSE contribution can be converted to a change of the Pt line resistance due to the magnetic field reversal. This geometric scaling approach is similar to prior extraction of a UMR in thin HM/FM bilayers [2], except that the magnetoresistive, anomalous, and planar Hall effects in the FM [15] are absent in the FI here. As shown in Fig. 3(a), the obtained UMR increases linearly with the current and exceeds $10^{-6}$.

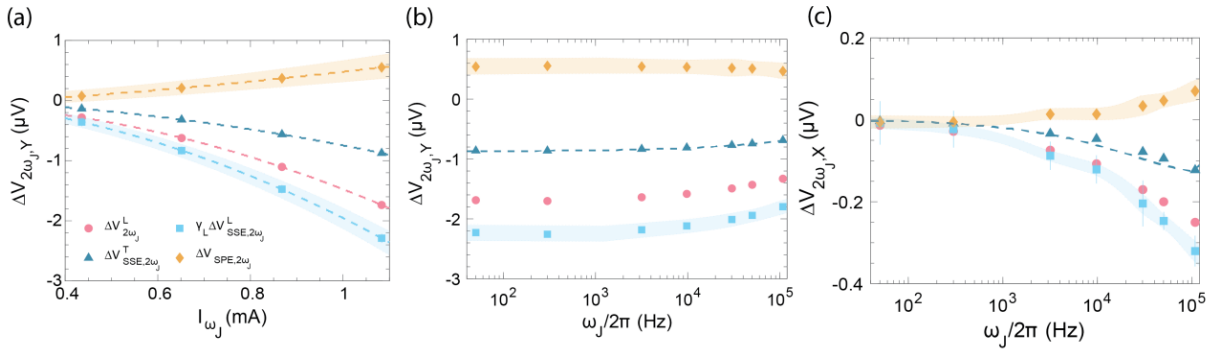

**FIG. 2.** Current and frequency dependence of SSE. (a) The measured change in $V_{2\omega_J,Y}$ (symbols) as a function of the rms amplitude of the current ($I$) in the Pt line as the field is swept from -20 mT to +20 mT at $\omega_J/2\pi = 50.17$ Hz. The lines are quadratic fits to the data. The out-

of-phase (b) and in-phase (c) components of the measured change in $V_{2\omega_J}$ (symbols) as a function of the frequency of the current at an rms amplitude of 1.08 mA in the Pt line as the field is swept from −40 mT to +40 mT. Each symbol represents the second harmonic component of the bridge voltage ($\Delta V_{2\omega_J}$, circles), transverse SSE signal ($\Delta V_{SSE,2\omega_J}^T$, triangles), longitudinal SSE contribution to $\Delta V_{2\omega_J}^L$ ($\gamma_L \Delta V_{SSE,2\omega_J}^L$, squares), and $\Delta V_{SPE,2\omega_J} = \Delta V_{2\omega_J}^L - \gamma_L \Delta V_{SSE,2\omega_J}^L$ (diamonds), where $\Delta V_{SSE,2\omega_J}^L = -(L_h/W_h)\Delta V_{SSE,2\omega_J}^T$ and $\gamma_L$ = -0.24 based on bridge circuit analysis. $\Delta V_{SSE,2\omega_J}^T$ are compared with the analytical solutions (dashed lines of (b) and (c)) with $\lambda = 830$ nm, $l_{mp} = 250$ nm, $\kappa_M = 0.1$ Wm$^{-1}$K$^{-1}$, $\sigma_m = 2.15 \times 10^6$ Sm$^{-1}$, $S_m = 63.5$ μVK$^{-1}$, and $S_s = 65.5$ μVK$^{-1}$. The error bars indicate the statistical uncertainty with 95% confidence and the shaded bands represent the additional systematic errors caused by the uncertainty in $L_h/w_H$.

The UMR calculated from prior theory [18] can increase by two orders of magnitude to our observed value only when the magnon gap in YIG is suppressed considerably to increase interface spin transport and its sensitivity to the sign of the magnon spin accumulation. Because the temperature is well below the Curie temperature and the magnetic field is finite in our measurement, we do not expect a large reduction of the magnon gap. On the other hand, several measurements of HM/FI heterostructures have found that the HM phonon temperature depends linearly on the current when **M** and **σ** are collinear [5,6]. The observation has been attributed to the SPE associated with the spin-mediated heat transport across the interface, which was not accounted for in the prior UMR theories [18]. A change in the phonon temperature is expected to alter both the phonon-electron scattering rate and the resistivity of the Pt. If the observed UMR is attributed to such an SPMR, the Pt phonon temperature change due to the SPE would be $\langle \vartheta_{p,Pt,\omega_J} \rangle$ = UMR/TCR, where TCR is the temperature coefficient of resistance of the Pt. A linear fit of the calculated low-frequency $\langle \vartheta_{p,Pt,\omega_J} \rangle$ with the charge current density ($j_c$) in the Pt is shown by the dashed line in Fig. 3(a) and yields a slope of $4.5 \times 10^{-13}$ K m$^2$ A$^{-1}$, which is close to previous SPE measurement results [6,27]. In addition, the sign of the calculated $\langle \vartheta_{p,Pt,\omega_J} \rangle$ is fully consistent with the SPE. As the $\Delta V_{2\omega_J}^L$ signals are not detectable in a Pt/GGG control sample (see Supplementary Material section G [25]), we rule out contributions from ordinary thermoelectric effects to the observed UMR in the Pt/YIG/GGG sample.

As $\omega_J/2\pi$ exceeds $10^4$ Hz in Fig. 3(b, c), the X and Y components of the UMR and $\langle\vartheta_{p,Pt,\omega_J}\rangle$ begin to vary with frequency. With the measured $\langle\vartheta_{p,Pt,\omega_J}\rangle$ and $\Delta V_{2\omega_J}^T$ normalized by the low-frequency limit $\langle\vartheta_{p,Pt,\omega_J\to 0,X}\rangle$ and $\Delta V_{2\omega_J\to 0,Y}^T$ measured at $\omega_J/2\pi = 50.17$ Hz, the proposed SPE signal shows a similar frequency dependence as the SSE (Fig. 4a). A recent two-temperature model [28] has used the frequency-dependence of $\Lambda$ to explain the reported SSE signal roll off at higher frequencies in thinner YIG films without either invoking a hypothesis based on the thickness confinement effect on the magnon dispersion [29] or accounting for the magnon chemical potential ($\mu_m$) [22,30]. Here, we establish a model to account for both $\mu_m$ and the magnon-phonon temperature difference ($\vartheta_{m-p}$) and obtain the following analytical solutions of their frequency-modulation components

$$\hat{\vartheta}_{m-p} = \sum_{j=0,1,2} e_j \left[a_j \exp\left(\frac{z}{l_j}\right) + b_j \exp\left(\frac{-z}{l_j}\right)\right] \tag{1}$$

$$\hat{\mu}_m = \sum_{j=0,1,2} d_j \left[a_j \exp\left(\frac{z}{l_j}\right) + b_j \exp\left(\frac{-z}{l_j}\right)\right] \tag{2}$$

where the $a_j$ and $b_j$ coefficients are obtained from the boundary conditions, and $e_j$ and $d_j$ are determined by the phonon and magnon properties including the magnon conductivity ($\sigma_m$) and bulk spin coefficient ($S_m$). In addition, $l_2$ approaches the smaller one between $l_{mp}$ and $\lambda$ when the two are very different and $S_m$ is small. In comparison, the other two length scales $l_0$ and $l_1$ follow $\Lambda$ and the larger one between $l_{mp}$ and $\lambda$, respectively, until an anti-crossing behavior emerges in Fig. 4(b) when $\Lambda$ is reduced to the latter. The anti-crossing behavior is accompanied with the emergence of wavelike profiles in both the temperatures and $\mu_m$ [Fig. S11], which results in kinks or bumps in the calculated frequency-dependent SPE and SSE signal [Fig. 4(a)]. As such features are absent in the frequency range of the measurement, λ and $l_{mp}$ of our samples should be within the Λ value of 1.5 µm at the second harmonic component of the $10^5$ Hz frequency range. This finding agrees with a prior report of a 1-µm scale magnon propagation length in SSE measurements [31].

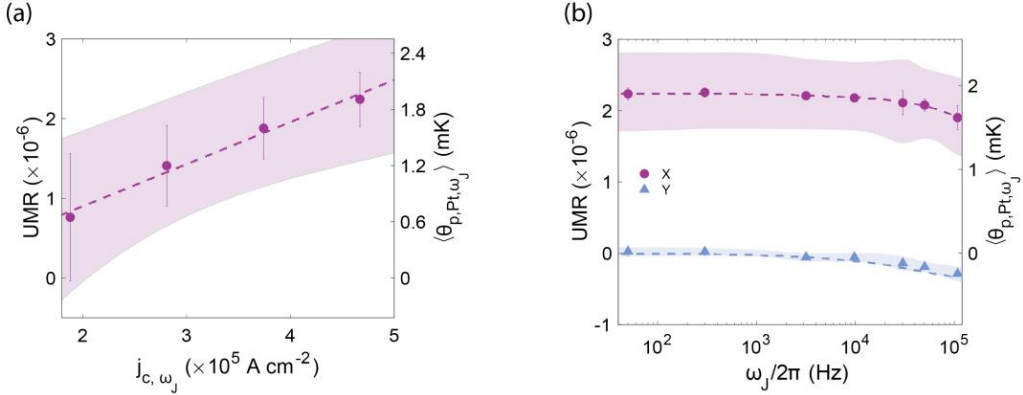

**FIG. 3**. Current and frequency dependence of unidirectional magnetoresistance. Unidirectional magnetoresistance (left axis) and equivalent Pt phonon temperature change (right axis) due to the SPE at $\varphi = -90°$ as a function of (a) the rms amplitude of the current density in the Pt at 50.17 Hz frequency and (b) the current frequency at $4.7 \times 10^5$ A cm$^{-2}$ current density. The dashed line of (a) is a linear fit of the measurement data (symbols) with a slope of $4.5 \times 10^{-13}$ K m$^2$ A$^{-1}$, while the dashed lines of (b) are obtained from the analytical solutions with $\lambda = 830$ nm, $l_{mp} = 250$ nm, $\kappa_M = 0.1$ Wm$^{-1}$K$^{-1}$, $\sigma_m = 2.15 \times 10^6$ Sm$^{-1}$, $S_m = 55.0$ µVK$^{-1}$, and $S_s = 168.0$ µVK$^{-1}$. The error-bars indicate the propagated uncertainty from the individual measurements and the shaded bands represent the additional systematic errors caused by uncertainty in $L_h/w_H$.

We find that our frequency-dependent SSE signal [Fig. 4(a)] can be fitted well with $\lambda = 0.83$ µm and $l_{mp} = 250$ nm [23,24], magnon thermal conductivity $\kappa_M = 0.1$ Wm$^{-1}$K$^{-1}$, $\sigma_m = 2.15 \times 10^6$ Sm$^{-1}$, $S_m = 63.5$ µVK$^{-1}$, and an interface spin Seebeck coefficient $S_s = 65.5$ µVK$^{-1}$. These and other properties used for the fitting are within one order of magnitude of prior reported theoretical and experimental estimates (see Supplementary Materials [25], Table S1). The same values can also fit the observed frequency-dependence of the SPE. Because the consideration of magnon chemical potential introduces three additional magnon transport properties ($\lambda$, $\sigma_m$, and $S_m$) in our model compared to the prior two-temperature model [28], we have not conducted an extensive search of optimum fitting values of the various magnon-mediated properties. In particular, the data can also be fitted with a smaller $l_{mp}$ below 10 nm based on earlier literature [21,22] together with a $\lambda$ in the 1 µm scale.

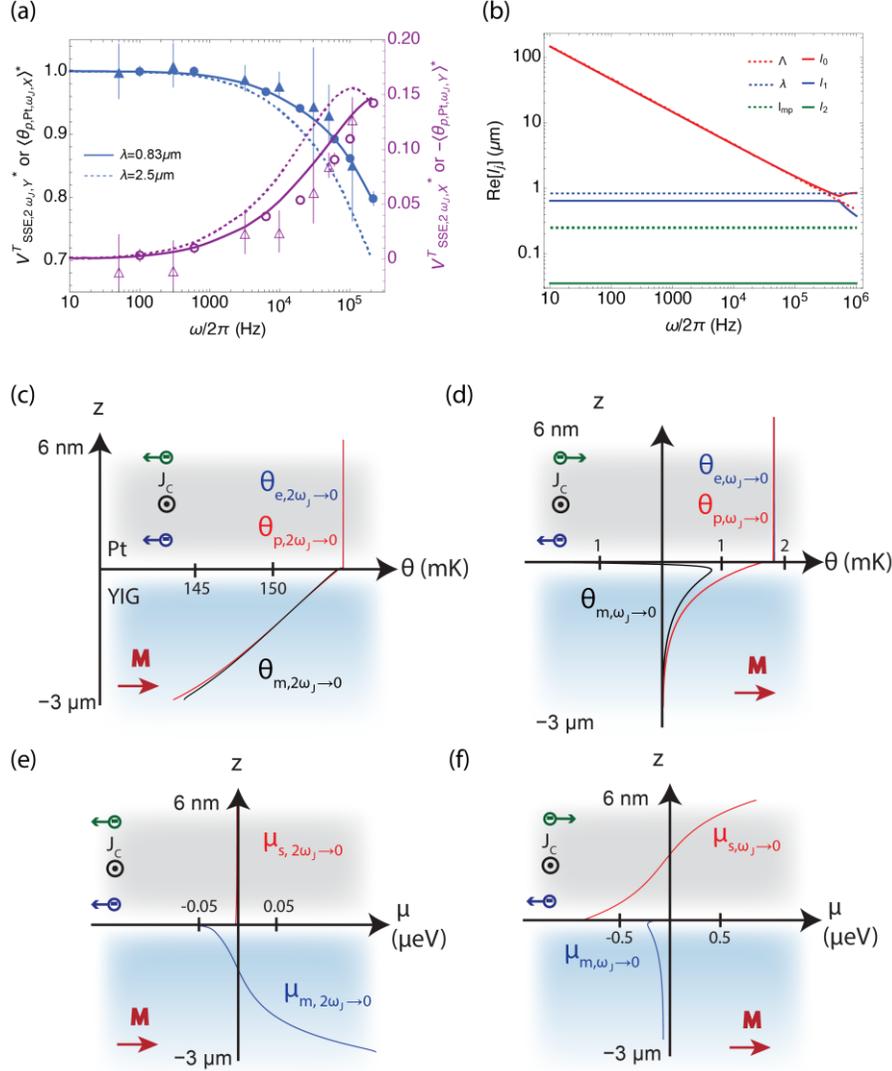

**FIG. 4.** (a) Comparison between the in-phase (filled symbols and blue lines) and out-of-phase (open symbols and purple lines) components of the normalized frequency-dependent second-harmonic SSE signals (circles) and first harmonic SPE signals (triangles) with the analytical solutions for the SSE (lines) at $l_{mp}$= 250 nm and $\lambda$ = 0.83 µm (solid lines) and $\lambda$ = 2.5 µm (dashed lines), respectively. The frequency $\omega$ equals $\omega_J$ for SPE and $2\omega_J$ for SSE. (b) Calculated $l_{j=0,1,2}$ at the $\omega = 2\omega_J$ frequency in comparison with $\Lambda$, $\lambda$ = 1 µm, and $l_{mp}$ = 250 nm. (c-f) Calculated second harmonic (c,e) and first harmonic (d,f) temperature and spin accumulation modulations for Pt

electrons ($\theta_e$, $\mu_s$), Pt and YIG phonons ($\theta_p$), and YIG magnons ($\theta_m$, $\mu_m$) at the low-frequency limit.

The analytical model yields a ratio $U_{sc} \equiv \frac{\Delta V^T_{2\omega_J \to 0, Y}}{\langle \vartheta_{p,Pt,\omega_J \to 0, X} \rangle}$ that is insensitive to most of the transport properties except for a weak dependence on the spin diffusion length ($\lambda_{Pt}$) in Pt (Fig. S12). This behavior reveals the Onsager reciprocity between the SSE and SPE coefficients. For $\lambda_{Pt} = 1.5$ nm, the calculated $U_{sc}$ is -183.6 µV/K, which is a factor of 2.5 larger than the measurement value and supports the presence of a much larger SPMR than the prior prediction of magnonic UMR effect [18]. One possible cause of the difference can be the geometric scaling approach that is used to evaluate the SSE effect in the longitudinal bridge signal along the long Pt line based on the transverse SSE measurement over its much smaller width. Due to the different magnon temperatures and chemical potentials (Fig. 4, c-f) involved in the second harmonic and first harmonic measurements, moreover, the thermal magnons responsible for the second harmonic SSE measurements are in a different frequency range compared to the magnons that are injected by the spin Hall effect to give rise to the first harmonic SPE signal, as discussed in a recent work [32]. As a result, the second-harmonic SSE and first-harmonic SPE fitting values can still differ because Onsager reciprocity for the SSE and SPE coefficients is only valid when the magnon temperatures, and similarly the chemical potentials, are at the same respective value in both effects (see Equations S104 and S91). Indeed, we find that a change of the fitting parameters to $S_m = 55$ µVK$^{-1}$, and $S_s = 168$ µVK$^{-1}$ can lead to satisfactory fitting of both the frequency-dependence and magnitude of the measured SPE signal. It should also be noted that spin-mediated interfacial heat transfer can also enable the transfer of Ettingshausen heat from Pt electrons to YIG magnons. However, the Ettingshausen effect does not produce the measured frequency dependence and can produce the magnitude of the observed low-frequency first harmonic signal only when Nernst coefficient is anomalously large (see Supplementary Materials section Q [25]).

These measurements and analyses show that the UMR in HM/FI heterostructures is much larger than prior theories that have only considered the interfacial spin flux but not the spin-mediated heat flux caused by electron-magnon scattering. The reversal of the magnetization collinear with the spin accumulation leads to a small magnitude change of the interface spin flux due to the variation of the magnon accumulation, but reverses the direction of the corresponding spin Peltier heat flux, thus producing a much larger UMR than expected from the spin flux alone. Hence, it is worth investigating whether the SPE also plays a role in the UMR observed in FM/NM metallic bilayers. By accounting for both magnon chemical potential and nonequilibrium magnon temperature, our analytical model shows that the frequency- and thickness-dependent SPE and SSE processes are caused by a reduction of either the thermal penetration depth or thickness to the fundamental magnon spin diffusion length and magnon-phonon thermalization length, which are within 1 μm for our samples. The frequency-dependent measurements and analysis introduced here offer a unique probe of the fundamental length scales of spin and energy carriers and facilitate the exploration of nonequilibrium magnons as carriers of information and energy.

**Acknowledgements**

We thank David Cahill for helpful discussions and for sharing the Pt/YIG interface thermal conductance measurement data. We additionally thank Anthony Stauzenberger, Uttam Ghoshal, Yoonho Seo, and Neal Hall for providing assistance in sputter deposition of thin Pt films. This work was supported by the US Army Research Office Multidisciplinary University Research Initiative (MURI) Grant W911NF-14-1-0016. One of the authors (HL) was supported by the US Office of Naval Research MURI Grant No. N00014-16-1-2436. S.E.S., A.W., and L.S. designed and carried out the low-frequency measurements. H.L., L.S., and S.E.S. designed and carried out the high-frequency measurements. L.S. developed the analytical models. S.E.S. and H.L. carried out calculations. All authors contribute to the writing of the manuscript. S.E.S. and H.L. contributed equally to this work.

Supplementary Materials

**Frequency-dependent Phonon-mediated Unidirectional Magnetoresistance in a Metal on an Insulator with Highly Nonequilibrium Magnons**

Sean Sullivan,[1,+,^] Hwijong Lee,[1,^] Annie Weathers,[2,#] Li Shi[1,2,*]

[1]Materials Science and Engineering Program, The University of Texas at Austin, Austin, Texas 78712

[2]Department of Mechanical Engineering, The University of Texas at Austin, Austin, Texas 78712

[+]Current address: Argonne National Laboratory, Lemont, Illinois 60439, USA

[#]Current address: MIT Lincoln Laboratory, Lexington, Massachusetts 02421, USA

[^]These authors contributed equally.

* Author to whom correspondence should be addressed: lishi@mail.utexas.edu

## A. Sample Fabrication

The samples were fabricated on 3 µm-thick, [111] commercially available (MTI) yttrium iron garnet (YIG) films grown on a gadolinium gallium garnet (GGG) substrate via liquid phase epitaxy. The YIG surface was cleaned in a piranha solution for 10 minutes and dried. Immediately after cleaning, 5.8 nm of Pt was deposited by radio-frequency (RF) sputtering. Standard photolithography and Ar plasma etching were used to pattern Hall bars with lateral dimensions 40 µm × 800 µm. Subsequently, 10 nm thick Ti and 70 nm thick Au were deposited through a photoresist mask via an electron beam deposition and subsequent liftoff process to form the bonding pads and leads. Figure S1 shows the photograph of one of the two Pt lines that were patterned on the YIG film with the same design and yielded similar measurement results. Additional Pt line control devices were patterned on [111] commercially-available GGG crystals and 10 µm-thick YIG on GGG. The YIG surface of the control samples was damaged by Ar plasma just prior to Pt sputtering.

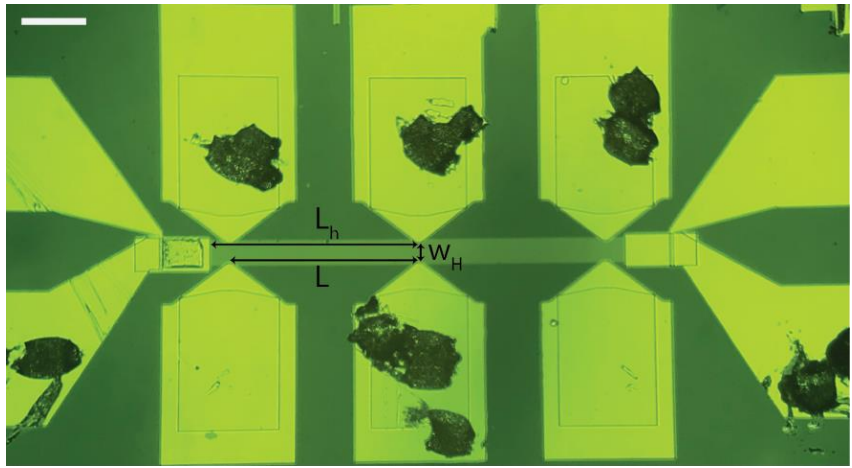

**FIG S1**. Optical image of the Pt/YIG sample. The white scale bar indicates 100 µm. $L_h$ is the half length of the Pt line. $L$ is the distance between the voltage probes, while $w_H$ is the distance between the two Hall probes. The black patches on the Au bonding pads are residue of the removed aluminum bonding wires.

## B. Spin Hall Magnetoresistance

All Measurements were carried out in a magnetic cryostat at room temperature. Figure S2 shows first harmonic ($1\omega_J$) transverse and longitudinal magnetoresistance data of Sample 1 that was measured with a sinusoidal excitation current in the Pt line at a frequency of $\nu_J = \omega_J/2\pi = 1017.1$ Hz. Root-mean-square (rms) amplitudes are used for all the modulated signals. For the longitudinal resistance measurement, we utilized a balanced circuit to nullify the large $1\omega_J$ component of the two voltage signals, $V_A$ and $V_B$, using two lock-in amplifiers. One arm of the circuit comprised the Pt/YIG heterostructure in a magnetic cryostat, while the other consisted of precision resistors and variable resistors outside the cryostat. Once both $V_A$ and $V_B$ were nullified, the analog outputs of the two amplifiers were measured in a differential configuration to obtain the longitudinal voltage as $V^L = V_A - V_B$ and to reject common mode noise. When the small imaginary component of the interfacial spin mixing conductance is ignored compared to its real component ($g_r$), the measured spin Hall magnetoresistance (SMR) can be expressed as [1–3]

$$MR^L \approx -\Theta_{SH}^2 \frac{\lambda_{Pt}}{l} \frac{\tanh^2 \frac{l}{2\lambda_{Pt}}}{\frac{\sigma_{xx}}{\lambda_{Pt} g_r} + 2\coth \frac{l}{\lambda_{Pt}}} \tag{S1}$$

where $\Theta_{SH} = 0.11$ is the spin Hall angle, $\lambda_{Pt} = 1.5$ nm is the reported spin diffusion length in Pt [3–5], $\sigma_{xx} = 1.85 \times 10^6$ Ohm$^{-1}$m$^{-1}$ is the measured longitudinal conductivity, and $l = 5.8$ nm is the thickness of the Pt film. Derivation of this equation is based on $m_y^2(H = 0) = \frac{1}{2}$, where $m_y$ is the *y* component of the unit vector (**m**) that is parallel to the magnetization (**M**) direction under an applied magnetic field (**H**). This equation is use to analyzed the measurement result to obtain a $g_r$ value of $1.11 \times 10^{14}$ $\Omega^{-1}$m$^{-2}$, which is comparable to the reported value for similar Pt/YIG interfaces [3]. When the electron spin polarization is collinear with **M**, the interface spin conductance ($g_s$) is given as [3]

$$g_s = \frac{3\zeta\left(\frac{3}{2}\right)}{2\pi s} \frac{g_r}{\Lambda_{dB}^3} \approx \Re g_r \tag{S2}$$

where *s* is the equilibrium spin density of the magnetic insulator, $\Lambda_{dB} = \sqrt{4\pi D_s/(k_B T)}$ is the magnon thermal de Broglie wavelength, $D_s = 8.4 \times 10^{-40}$ $Jm^{-2}$ is the spin stiffness [6], and $\Re$ is a constant which varies from 0.06 to 0.16 in literature.

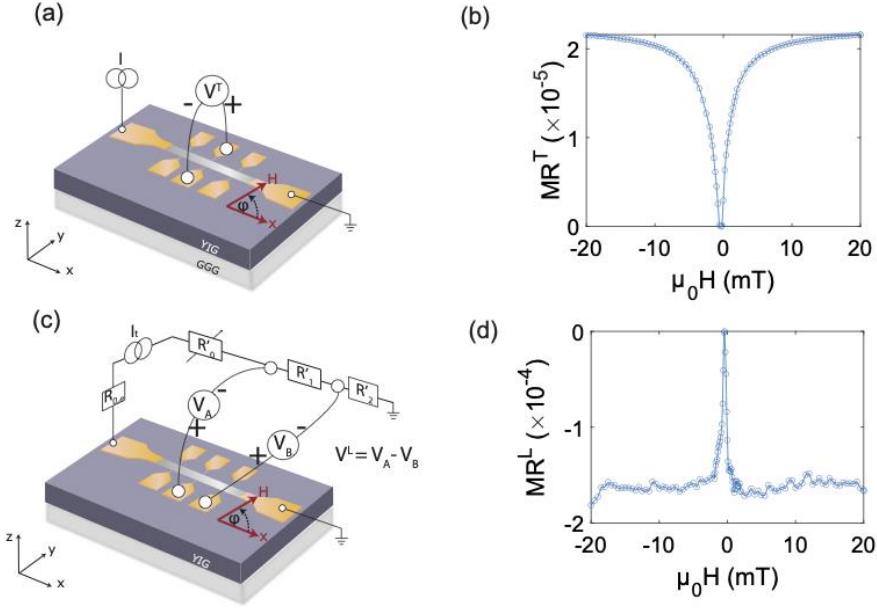

**FIG S2**. Transverse and longitudinal spin Hall magnetoresistance (SMR) measurements of Pt/YIG Sample 1. (a) Schematic illustration of the transverse lock-in electrical measurement configuration. (b) Magnetoresistance (MR) obtained from the in-phase (X), first harmonic frequency ($\nu_J = \omega_J/2\pi = 1017.1$ Hz) component of the voltage between the two middle transverse voltage probes as a function of the applied **H** that is collinear with the direction of the charge current, $\varphi = -90°$. Here, $\mu_0$ is the vacuum permeability and $MR^T \equiv \frac{\rho_{xy}(\mathbf{H}) - \rho_{xy}(\mathbf{H}=0)}{\rho_{xx}(\mathbf{H}=0)}$, where $\rho_{xx}$ and $\rho_{xy}$ are the longitudinal and transverse resistivity. (c) Longitudinal measurement configuration, where a balanced circuit is used to nullify the first harmonic component of the two bridge voltages, $V_A$ and $V_B$ at -20 mT field. (d) The longitudinal magnetoresistance, $MR^L \equiv \frac{\rho_{xx}(\mathbf{H}) - \rho_{xx}(\mathbf{H}=0)}{\rho_{xx}(\mathbf{H}=0)}$, measured as a function of **H** applied in the *xy*-plane with $\varphi = -90°$. The Pt line resistance ($R$) is obtained as the ratio between the first harmonic, rms amplitude ($V_{\omega_J}$) of the voltage drop between the two voltage probes on the Pt line and the rms amplitude ($I_{\omega_J}$) of the current flow in the Pt line. The rms amplitude of the sinusoidal current ($I_t$) from a current source was $I_0 = 2$ mA, about half of which flows in the Pt line. Based on a bridge circuit analysis, $MR^L = \frac{V_{\omega_J}(\mathbf{H}) - V_{\omega_J}(\mathbf{H}=0)}{\beta I_0 R(\mathbf{H}=0)}$, where $\beta = \frac{R'_t}{R_t(\mathbf{H}) + R'_t}\left[1 - \frac{L_t}{L}\frac{R_1(\mathbf{H}=0) + R_1'}{R_t(\mathbf{H}=0) + R'_t}\right]$, $R_t$ and $R'_t$ are the total resistances of the two arms of the bridge circuit made of Pt line and the precision resistors, respectively, and $L_t$ and $L$ are the total length and the segment length between the voltage probes of the Pt line.

## C. Charge and Spin Transport in Pt with non-uniform Temperatures

With the presence of a non-negligible spin Hall angle $\Theta_{SH}$, the electric current density component ($J_i$) along the $x_i$ axis of a Cartesian coordinate and the spin polarization current density tensor ($q_{ij}$), where the flow direction is along $x_i$ and the spin polarization direction is along $x_j$, are given by Dyakonov as [7]

$$J_i = J_i^0 + e\Theta_{SH}\epsilon_{ijk}q_{jk}^0 \tag{S3}$$

and

$$q_{ij} = q_{ij}^0 - \frac{\Theta_{SH}\epsilon_{ijk}}{e}J_k^0 \tag{S4}$$

where a sign change has been made to adopt the spin Hall angle sign definition of latter works [8], $e$ is the proton charge, $\epsilon_{ijk}$ is the unit antisymmetric tensor with the only non-zero components being $\epsilon_{123} = -\epsilon_{213} = \epsilon_{231} = -\epsilon_{321} = \epsilon_{312} = -\epsilon_{132} = 1$, and thermoelectric effects have been included in the expressions of

$$J_i^0 = -\sigma_{ij}\frac{\partial \Phi}{\partial x_j} - \sigma_{ik}S_{kj}\frac{\partial T}{\partial x_j} = \sigma_{ij}E_j + eD_{ij}\left(\frac{\partial n}{\partial x_j} - \frac{\partial n}{\partial T}\frac{\partial T}{\partial x_j}\right) - \sigma_{ij}S_{jk}\frac{\partial T}{\partial x_k} \tag{S5}$$

and

$$q_{ij}^0 = -\frac{P_j\sigma_{ik}E_k}{ne} - D_{ik}\left(\frac{\partial P_j}{\partial x_k} - \frac{P_j\partial n}{n\partial T}\frac{\partial T}{\partial x_k}\right) + \frac{P_j\sigma_{il}S_{lk}}{ne}\frac{\partial T}{\partial x_k} \tag{S6}$$

Here, $\Phi$ is the electrochemical potential, $\frac{\partial \Phi}{\partial x_j} = -E_j - \frac{1}{e}\frac{\partial \mu_c}{\partial x_j}$, $E_j$ is the electric field ($\boldsymbol{E}$) component along $x_j$, $\mu_c$ is the chemical potential, $\sigma_{ij}$ is the electrical conductivity tensor component, $S_{kj}$ is the thermoelectric coefficient tensor component, $T$ is the temperature, $n$ is the electron density, $\frac{\partial n}{\partial x_j} = \frac{\partial n}{\partial \mu_c}\frac{\partial \mu_c}{\partial x_j} + \frac{\partial n}{\partial T}\frac{\partial T}{\partial x_j}$, $P_j$ is the $x_j$ component of the electron spin polarization density vector ($\boldsymbol{P}$), and the diffusivity tensor component is defined as $D_{ij} \equiv \frac{1}{e}\left[\frac{\partial J_i^0}{\partial\left(\frac{\partial n}{\partial x_j}\right)}\right]_{\boldsymbol{E}=0,\boldsymbol{\nabla}T=0} = \frac{\sigma_{ij}}{e^2\frac{\partial n}{\partial \mu_c}}$.

Because $n$ and $\mu_c$ are constant in a metal such as Pt, the explicit expressions are reduced to

$$J_i = \sigma_{ij}\left(E_j - S_{jk}\frac{\partial T}{\partial x_k}\right) - e\Theta_{SH}\epsilon_{ijk}\left[D_{jl}\frac{\partial P_k}{\partial x_l} + \frac{P_k\sigma_{jl}}{ne}\left(E_l - S_{lm}\frac{\partial T}{\partial x_m}\right)\right] \tag{S7}$$

and

$$q_{ij} = -\frac{P_j\sigma_{ik}}{ne}\left(E_k - S_{kl}\frac{\partial T}{\partial x_l}\right) - D_{ik}\frac{\partial P_j}{\partial x_k} - \frac{\Theta_{SH}\epsilon_{ijk}}{e}\sigma_{kl}\left(E_l - S_{lm}\frac{\partial T}{\partial x_m}\right) \tag{S8}$$

The steady-state spin diffusion equation is expressed as

$$P_j = -\tau_s\frac{\partial q_{ij}}{\partial x_i} \tag{S9}$$

where $\tau_s$ is the spin relaxation time. As shown in Fig. S3, where $x = x_1, y = x_2, z = x_3$, the electrical current ($I$) flows in the Pt film along the $x_1$ direction, with $J_3 = 0$. Due to the large dimension along the $x_1$ and $x_2$ directions,

$$\frac{\partial E_2}{\partial x_3} = -\frac{\partial^2 \Phi}{\partial x_2 x_3} = \frac{\partial E_3}{\partial x_2} = 0 \tag{S10}$$

and

$$\frac{\partial E_1}{\partial x_3} = -\frac{\partial^2 \Phi}{\partial x_1 x_3} = \frac{\partial E_3}{\partial x_1} = 0 \tag{S11}$$

When **H** is colinear with the $x_l$ direction, the off-diagonal components of the conductivity, diffusivity and thermoelectric tensors vanishes when one of the subscripts is along $x_l$. In addition, $\sigma_{ij} = -\sigma_{ji}$ for $i \neq j$.

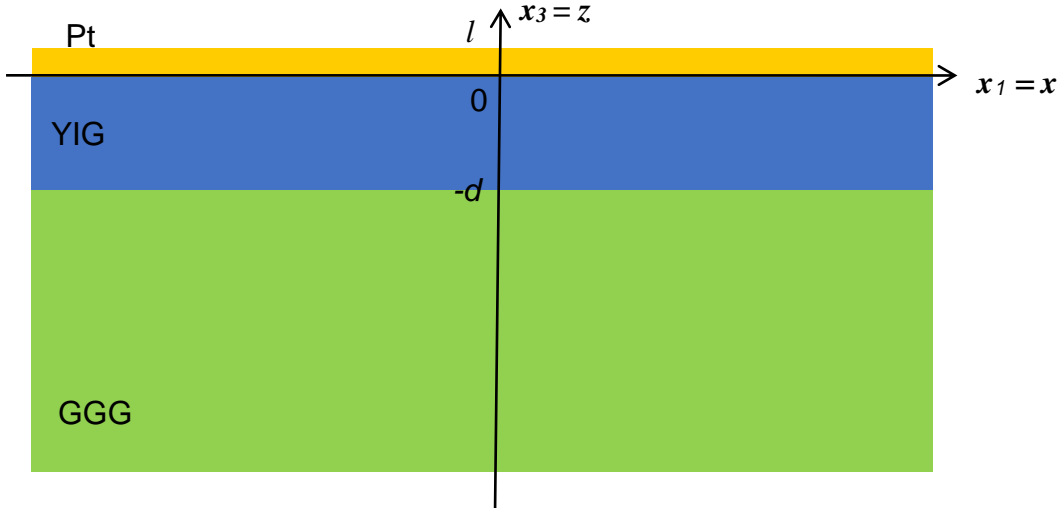

**FIG. S3**. Schematic illustration of the Pt/YIG/GGG structure, where $x_2 = y$ points normal into the page for a right-handed $x_1$ $x_2$ $x_3$ Cartesian coordinate and $x_1 = x$ and $x_3 = z$.

## D. Longitudinal Electric Fields and Spin Accumulation in the Pt Line

Equations S7-S9 can be expanded into a total of nine equations for $J_1, J_2, J_3, q_{31}, q_{32}, q_{33}, P_1, P_2,$ and $P_3$ for the case that the current is colinear with $x_1$ and $\mathbf{H}$ is colinear with $x_2$. Because the spin Hall effect and the YIG magnetization produce spin polarization colinear with the $x_2$ direction, a solution of vanishing $P_1, P_3, q_{31}, q_{33}, E_2,$ and $J_2$ satisfy the spin diffusion equations for the $P_1$ and $P_3$ components and the boundary conditions. The $E_1$ and $E_3$ terms are substituted from the set of equations and the high-order terms are ignored to obtain

$$J_1 = \sigma_{11}\left(E_1 - S_{13}\frac{\partial T}{\partial x_3}\right) + eD_{33}\Theta_{SH}\frac{\partial P_2}{\partial x_3} \tag{S12}$$

and

$$q_{32} = -(1 + \Theta_{SH}^2)D_{33}\frac{\partial P_2}{\partial x_3} + \frac{\Theta_{SH}}{e}J_1 \tag{S13}$$

Taking partial derivative of Equation S12 over $x_3$ leads to

$$\frac{\partial J_1}{\partial x_3} = -\sigma_{11}S_{13}\frac{\partial^2 T}{\partial x_3^2} + e\Theta_{SH}D_{33}\frac{\partial^2 P_2}{\partial x_3^2} \tag{S14}$$

After using Equation S13 to substitute the $q_{32}$ terms in the spin diffusion equation for the $P_2$ component, we obtain

$$-D_{33}\frac{\partial^2 P_2}{\partial x_3^2} - \frac{\Theta_{SH}}{e}\sigma_{11}S_{13}\frac{\partial^2 T}{\partial x_3^2} + \frac{P_2}{\tau_s} = 0 \tag{S15}$$

When the thermal non-equilibrium between electrons and phonons in Pt is ignored, the heat diffusion equation in Pt gives

$$\frac{\partial^2 T}{\partial x_3^2} = -\frac{J_1^2}{\sigma_{11}\kappa_{Pt}} \approx -\frac{I^2}{\sigma_{11}\kappa_{Pt}W^2 l^2} \tag{S16}$$

and

$$\frac{\partial T}{\partial x_3} \approx \left(\frac{\partial T}{\partial x_3}\right)_c - \frac{I^2}{\sigma_{11}\kappa_{Pt}W^2 l^2}x_3 \tag{S17}$$

where $\left(\frac{\partial T}{\partial x_3}\right)_c$ is a constant, $\kappa_{Pt}$, $W$, and $l$ are thermal conductivity, width, and thickness of the Pt line. These results are used to expand the spin diffusion equation as

$$-D_{33}\frac{\partial^2 P_2}{\partial x_3^2} + \frac{P_2}{\tau_s} = -\frac{\Theta_{SH}S_{13}I^2}{e\kappa_{Pt}W^2 l^2} \tag{S18}$$

The solution is

$$P_2 = A\cosh\left(\frac{x_3}{\lambda_{Pt}}\right) + B\sinh\left(\frac{x_3}{\lambda_{Pt}}\right) - \frac{\tau_s \Theta_{SH} S_{13} I^2}{e\kappa_{Pt} W^2 l^2} \tag{S19}$$

where the spin flip diffusion length in Pt is defined as

$$\lambda_{Pt} = \sqrt{D_{33}\tau_s} \tag{S20}$$

This solution is used to obtain

$$J_1 = \frac{e\Theta_{SH} D_{33}}{\lambda_{Pt}}\left[A\sinh\left(\frac{x_3}{\lambda_{Pt}}\right) + B\cosh\left(\frac{x_3}{\lambda_{Pt}}\right)\right] - \frac{S_{13} I^2}{\kappa_{Pt} W^2 l^2} x_3 + C \tag{S21}$$

and

$$q_{32} = -\frac{D_{33}}{\lambda_{Pt}}\left[A\sinh\left(\frac{x_3}{\lambda_{Pt}}\right) + B\cosh\left(\frac{x_3}{\lambda_{Pt}}\right)\right] - \frac{\Theta_{SH} S_{13} I^2}{e\kappa_{Pt} W^2 l^2} x_3 + \frac{\Theta_{SH}}{e} C \tag{S22}$$

The three constants $A$, $B$, and $C$ are determined from the boundary conditions. The current in the Pt line is

$$I = \int_0^l J_1 W dx_3 = We\Theta_{SH} D_{33}\left\{A\left[\cosh\left(\frac{l}{\lambda_{Pt}}\right) - 1\right] + B\sinh\left(\frac{l}{\lambda_{Pt}}\right)\right\} - \frac{S_{13} I^2}{2\kappa_{Pt} W} + WlC \tag{S23}$$

At the top surface of the Pt film,

$$q_{32}(x_3 = l) = -\frac{D_{33}}{\lambda_{Pt}}\left[A\sinh\left(\frac{l}{\lambda_{Pt}}\right) + B\cosh\left(\frac{l}{\lambda_{Pt}}\right)\right] - \frac{\Theta_{SH} S_{13} I^2}{e\kappa_{Pt} W^2 l} + \frac{\Theta_{SH}}{e} C = 0 \tag{S24}$$

In addition, $P_2 = n_+ - n_-$, where $n_+$ and $n_-$ are the number densities of the spin-up and spin-down electrons in Pt with the spin aligned parallel and anti-parallel to the $x_2$ direction. Meanwhile, $n_+$ and $n_-$ depend on their respective electrochemical potential energies, $\mu_+$ and $\mu_-$, with the average $\mu_e \equiv \frac{\mu_+ + \mu_-}{2} = -e\Phi$. The spin flux across the Pt/YIG interface ($x_3 = 0$) is calculated as [9]

$$J_s^i \equiv -eq_{32}(x_3 = 0) = m_2 g_s\left(\frac{\mu_s^i - \mu_m^i}{e} + S_s \vartheta_{m-e}^i\right) = \frac{eD_{33}}{\lambda_{Pt}} B - \Theta_{SH} C \tag{S25}$$

where $m_2$ is the component of **m** along the $+x_2$ direction and takes the value of 1 and -1 when **M** is parallel and anti-parallel to $+x_2$, $\mu_s \equiv m_2(\mu_+ - \mu_-) \equiv -m_2 eV_s$ is the electron spin accumulation, $\mu_m$ is the magnon spin accumulation or magnon chemical potential, the superscript $i$ is used to denote the value at the Pt/YIG interface, and $S_s \equiv -\frac{1}{e}\left[\frac{\partial(\mu_s^i - \mu_m^i)}{\partial \vartheta_{m-e}^i}\right]_{J_s^i=0}$ is the interface spin Seebeck coefficient, with $L_s \equiv g_s S_s$.

The boundary conditions are used to obtain

$$A = \frac{-\tanh\left(\frac{l}{2\lambda_{Pt}}\right)\frac{\Theta_{SH}I}{Wl} - \left[\coth\left(\frac{l}{\lambda_{Pt}}\right) + \frac{\Theta_{SH}^2\lambda_{Pt}}{l}\right]J_s^i}{\frac{eD_{33}}{\lambda_{Pt}}\left[1 + \frac{2\Theta_{SH}^2\lambda_{Pt}}{l}\tanh\left(\frac{l}{2\lambda_{Pt}}\right)\right]} - \frac{I^2 S_{13}\Theta_{SH}\lambda_{Pt}\left\{\frac{1+\cosh\left(\frac{l}{\lambda_{Pt}}\right)}{\sinh\left(\frac{l}{\lambda_{Pt}}\right)} + \frac{2\Theta_{SH}^2\lambda_{Pt}}{l}\right\}}{2eD_{33}\kappa_{Pt}lW^2\left(1 + \frac{2\Theta_{SH}^2\lambda_{Pt}}{l}\tanh\left(\frac{l}{2\lambda_{Pt}}\right)\right)} \quad (S26)$$

$$B = \frac{\left[1 + \frac{\Theta_{SH}^2\lambda_{Pt}}{l}\tanh\left(\frac{l}{2\lambda_{Pt}}\right)\right]J_s^i + \frac{\Theta_{SH}I}{Wl}}{\frac{eD_{33}}{\lambda_{Pt}}\left[1 + \frac{2\Theta_{SH}^2\lambda_{Pt}}{l}\tanh\left(\frac{l}{2\lambda_{Pt}}\right)\right]} + \frac{S_{13}\Theta_{SH}\lambda_{Pt}I^2}{2eD_{33}\kappa_{Pt}W^2l} \quad (S27)$$

and

$$C = \frac{-\frac{\Theta_{SH}\lambda_{Pt}}{l}\tanh\left(\frac{l}{2\lambda_{Pt}}\right)J_s^i + \frac{I}{Wl}}{1 + \frac{2\Theta_{SH}^2\lambda_{Pt}}{l}\tanh\left(\frac{l}{2\lambda_{Pt}}\right)} + \frac{S_{13}I^2}{2\kappa_{Pt}W^2l} \quad (S28)$$

For a small $P_2$

$$\mu_s = \frac{\partial \mu_s}{\partial P_2}P_2 = m_2 \frac{2e^2 D_{33}}{\sigma_{33}}P_2 \quad (S29)$$

At the interface,

$$\mu_s^i = -\psi\left(eS_s\vartheta_{m-e}^i - \mu_m^i\right) - m_2 e\varsigma\eta \frac{2\lambda_{Pt}}{\sigma_{33}}\frac{\Theta_{SH}S_{13}I^2}{2\kappa_{Pt}W^2l} - m_2 e\eta \frac{2\lambda_{Pt}}{\sigma_{33}}\frac{\Theta_{SH}I}{Wl}\tanh\left(\frac{l}{2\lambda_{Pt}}\right) \quad (S30)$$

where

$$\psi \equiv g_s\eta \frac{2\lambda_{Pt}}{\sigma_{33}}\left[\coth\left(\frac{l}{\lambda_{Pt}}\right) + \frac{\Theta_{SH}^2\lambda_{Pt}}{l}\right], \quad (S31)$$

the backflow correction factor is

$$\eta = \left\{1 + \frac{2g_s\lambda_{Pt}}{\sigma_{33}}\coth\left(\frac{l}{\lambda_{Pt}}\right) + \frac{2\Theta_{SH}^2 g_s\lambda_{Pt}^2}{\sigma_{33}l}\left[1 + \frac{\sigma_{33}}{g_s\lambda_{Pt}}\tanh\left(\frac{l}{2\lambda_{Pt}}\right)\right]\right\}^{-1} \quad (S32)$$

and

$$\varsigma = \left(1 + \frac{2\Theta_{SH}^2\lambda_{Pt}}{l}\tanh\left(\frac{l}{2\lambda_{Pt}}\right)\right)\left(\tanh\left(\frac{l}{2\lambda_{Pt}}\right) + \frac{2\lambda_{Pt}}{l}\right) \quad (S33)$$

This $\eta$ result is reduced to the expression in a prior report [5] when the $\Theta_{SH}^2$, $\theta_{m-e}$ and $S_{13}$ terms are ignored, and the driving force is taken as $V_s^i/2$ in [7] instead of just $V_s^i$ in [9].

The longitudinal electric field for orthogonal in-plane $H$ and $I$ is obtained from the $J_1$ equation as

$$E_1 = \frac{C}{\sigma_{11}} + S_{13}\left(\frac{\partial T}{\partial x_3}\right)_c \quad (S34)$$

which is expressed as

$$E_1 = -\frac{\partial \Phi}{\partial x_1} = E_\Omega^L + E_N^L + E_{SSE}^L \quad (S35)$$

where the longitudinal electric fields caused by the Ohmic ($\Omega$) loss, Nernst (N) effect, and interface spin Seebeck effect (SSE) for $\mathbf{H}$ colinear with $x_2$ and $I$ colinear with $x_1$ are

$$E_\Omega^L = \frac{1 + \frac{g_s \Theta_{SH}^2 \lambda_{Pt}^2}{\sigma_{33}\eta l}\tanh^2\left(\frac{l}{2\lambda_{Pt}}\right)}{1 + \frac{2\Theta_{SH}^2 \lambda_{Pt}}{l}\tanh\left(\frac{l}{2\lambda_{Pt}}\right)} \frac{I}{\sigma_{11} W l} \quad (S36)$$

$$E_N^L = \left[1 + \frac{\frac{g_s \eta \varsigma I^2}{\sigma_{11}\kappa_{Pt}W^2 l}\frac{\Theta_{SH}^2 \lambda_{Pt}^2}{\sigma_{33} l}\tanh\left(\frac{l}{2\lambda_{Pt}}\right)}{\left\{\frac{I^2}{2\sigma_{11}\kappa_{Pt}W^2 l} + \left(\frac{\partial T}{\partial x_3}\right)_c\right\}\left\{1 + \frac{2\Theta_{SH}^2 \lambda_{Pt}}{l}\tanh\left(\frac{l}{2\lambda_{Pt}}\right)\right\}}\right] S_{13}\left[\left(\frac{\partial T}{\partial x_3}\right)_c + \frac{I^2}{2\sigma_{11}\kappa_{Pt}W^2 l}\right] \quad (S37)$$

and

$$E_{SSE}^L = -g_s \frac{\eta \Theta_{SH} \lambda_{Pt}}{\sigma_{11} l}\tanh\left(\frac{l}{2\lambda_{Pt}}\right) m_2 \left(S_s \vartheta_{m-e}^i - \frac{\mu_m^i}{e}\right) \quad (S38)$$

The last equation is reduced to the SSE expression in a prior report [10] when the $\mu_m^i$ term, the small $\Theta_{SH}^2$ term, and the other high-order terms are ignored.

### E. Spin Seebeck Effect

The SSE signal at the $2\omega_J$ frequency is caused by the heat flux induced by Joule heating in the Pt. When the magnetic field is rotated clockwise by $90°$ from the longitudinal measurement at $\varphi = -90°$ to the transverse measurements at $\varphi = 180°$, as shown in Fig. 1(a,d),

$$E_{SSE,2\omega_J}^L(\varphi = -90°) = -E_{SSE,2\omega_J}^T(\varphi = 180°) = \frac{V_{SSE,2\omega_J}^T(\varphi = 180°)}{w_H} \quad (S39)$$

where $E_{SSE,2\omega_J}^T(\varphi = 180°)$ is the SSE contribution to $E_2$ for the transverse measurement, and $V_{SSE,2\omega_J}^T(\varphi = 180°)$ is the corresponding measured second-harmonic transverse SSE voltage, and $w_H$ is the gap between the two Hall probes along the $x_2$ direction. A similar relationship is expected for the DC component caused by the Joule heating.

A Zurich Instruments ultrahigh frequency lock-in (UHFLI) amplifier was used in the measurement of $V_{SSE,2\omega_J}^T(\varphi = 180°)$, where the constant voltage output of the UHFLI ($V_L$) supplied an electrical current in the Pt film [Fig. 1(a)]. The voltage drops between the transverse probes were measured while the field was swept between -40 and +40 mT and $V_L$ was fixed at 1.0 V. The range of the charge current frequency $v_J$ ranged from 50.17 Hz to 108 kHz. The results of 50 to 1200 measurements are averaged to reduce signal-to-noise and calculate the statistical error in the signal change ($\Delta V$) when the magnetic field was changed from -40 mT to 40 mT.

For the measurement data shown in Fig. 1, the electrical current in the Pt line is modulated at a low angular frequency ($\omega_J$)

$$I = \sqrt{2}I_{\omega_J}\sin\omega_J t \quad (S40)$$

The resulted Joule heating are proportional to $I^2$ and modulated according to $\sin^2\omega_J t = \frac{1-\cos 2\omega_J t}{2}$, which contains a negative second-harmonic out-of-phase (Y) component. As such, $\vartheta_{m-e,2\omega_J,Y}^i$, the Y component of $\vartheta_{m-e,2\omega_J}^i$, is positive. For a positive $\vartheta_{m-e,2\omega_J,Y}^i$ and a positive $H = |\mathbf{H}|$ value at $\varphi = 180°$, the magnon spin polarization is antiparallel to $x_1$ and $q_{31,2\omega_J,Y}^0$ at the interface is negative. As $\Theta_{SH}$ is positive for Pt, the $e\Theta_{SH}\epsilon_{231}q_{31,2\omega_J,Y}^0$ term in the $J_2$ expression of Equation S3 is negative and antiparallel to the $x_2$ direction, resulting in negative charge accumulated at the positive terminal of the voltmeter in Fig. 1(a) and thus contributing to a negative value to the measured $V_{2\omega_J,Y}^T(\varphi = 180°)$, in agreement with the measurement results shown in Fig. 1(b). In addition, a negative $S_s$ in Equation S38 yields the observed $V_{2\omega_J,Y}^T(\varphi = 180°)$ sign.

### F. Spin Peltier Effect

The interface spin accumulation in the Pt annihilates or injects magnons. The nonequilibrium magnons are thermalized with the phonons in YIG over the length $l_{mp}$ and alters the YIG phonon temperature distribution. This spin-mediated heat transfer across interface changes the electron temperature, which is thermalized with the Pt phonon temperature over another thermalization length $\ell_{ep}$. At low frequencies, this spin Peltier effect (SPE) causes a first-harmonic, in-phase modulation ($\sqrt{2}\langle\vartheta_{p,Pt,\omega_{J,X}}\rangle\sin\omega_J t$) of the Pt phonon temperature. Due to electron-phonon scattering that is responsible for the measured temperature coefficient of resistance (TCR) in Fig.

S4, a first-harmonic modulation appears in the resistance, $R_{Pt} = \frac{L_{Pt}}{\sigma_{xx}Wl}$ of the Pt line with length $L_{Pt}$. The first-harmonic SPE modulation ($\sqrt{2}R_{Pt,\omega_J,X}\sin\omega_J t = \sqrt{2}R_{Pt}\text{TCR}\langle\vartheta_{p,Pt,\omega_J}\rangle\sin\omega_J t$) of the Pt line resistance mixes with the supplied current to produce a second-harmonic modulation of the $E_\Omega^L$ term in Equation S36, resulting in a second harmonic Y component in the longitudinal voltage $V^L$ along the Pt line at low frequencies as

$$V_{SPE,2\omega_{J,Y}}^L(\varphi = -90°) = -\frac{1 + \frac{g_s\Theta_{SH}^2\lambda_{Pt}^2}{\sigma_{33}\eta l}\tanh^2\left(\frac{l}{2\lambda_{Pt}}\right)}{1 + \frac{2\Theta_{SH}^2\lambda_{Pt}}{l}\tanh\left(\frac{l}{2\lambda_{Pt}}\right)} I_{\omega_J} R_{Pt,\omega_J,X} \quad (S41)$$

When **M** is rotated to be collinear with the current and thus orthogonal to the polarization of the electron spin accumulation at the interface, a spin transfer torque is applied at the interface without transmitting the spin current and the associated energy, eliminating the SPE [5,11]. Thus, the SPE effect is absent in the transverse Hall measurement geometry with collinear **M** and **I**.

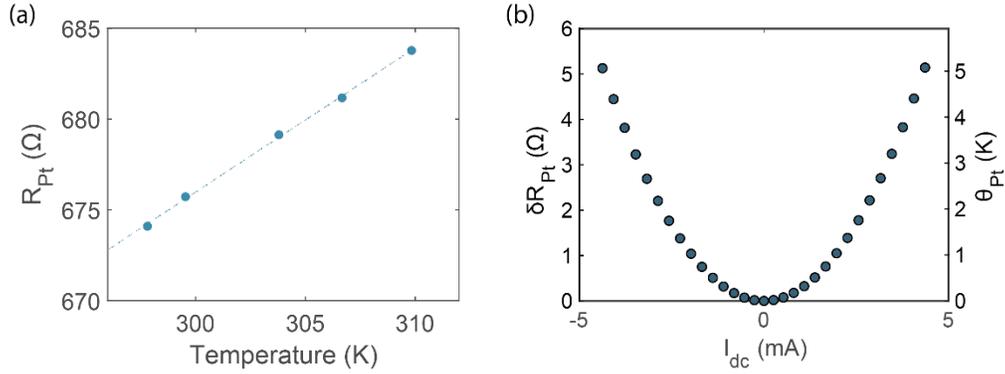

**FIG. S4**. Temperature coefficient of resistance (TCR) and DC electrothermal measurements of the Pt/YIG samples. (a) Four-probe resistance (R) of Pt line sample 1 as function of the sample stage temperature (T). The dashed line is a linear fit, which is used to obtain the temperature coefficient of resistance $\text{TCR} = \frac{dR}{R_{300K}dT} = 1.2 \times 10^{-3}$ K$^{-1}$, where $R_{300K}$ is the R value measured at 300 K. (b) The measured resistance change ($\delta R_{Pt}$) and temperature rise, $\theta_{Pt} = \delta R_{Pt}/(R_{Pt}\text{TCR})$, as a function of the DC current in Pt line sample 1.

### G. Conventional Thermoelectric Effects

The diagonal and off-diagonal components in the conventional thermoelectric tensor are caused by Seebeck and Nernst effects, respectively. Because the temperature gradients are quadratic functions of the electrical current, the conventional thermoelectric effects result in a contribution to the second harmonic components, $V^L_{2\omega_J}$ and $V^T_{2\omega_J}$, of the measured longitudinal bridge and transverse Hall voltages. To evaluate these thermoelectric contributions, we have performed additional measurements on a Pt/GGG and another Pt/YIG/GGG sample where the YIG sample was damaged by Ar$^+$ ion sputtering prior to deposition of Pt on the YIG. The interface spin-mixing conductance is expected to vanish for the Pt/GGG sample because GGG is not magnetically ordered, and decreased for the damaged YIG sample compared to the Pt/YIG where Ar$^+$ ions were not used to damage the YIG surface prior to Pt deposition. In these control experiments, the bridge circuit was made of two parallel Pt lines with antiparallel currents, as shown in Fig. S5(a), in order to double the $\Delta V^L_{2\omega,Y}$ signal caused by the SSE and SPE compared to the single Pt line configuration. At a low frequency where the second harmonic response is expected in the out-of-phase components, the obtained second harmonic, out-of-phase longitudinal bridge signal, $\Delta V^L_{2\omega_J,Y}(\varphi = -90°)$, depicted in Fig. S5(b) vanishes for the Pt/GGG sample, and decreases considerably for the damaged YIG sample compared to the values shown in Fig. 2(a). The measurement results for the Pt/GGG sample indicates that the conventional thermoelectric effects are not detectable in the measured $\Delta V^L_{2\omega_J,Y}$ caused by the magnetic field reversal.

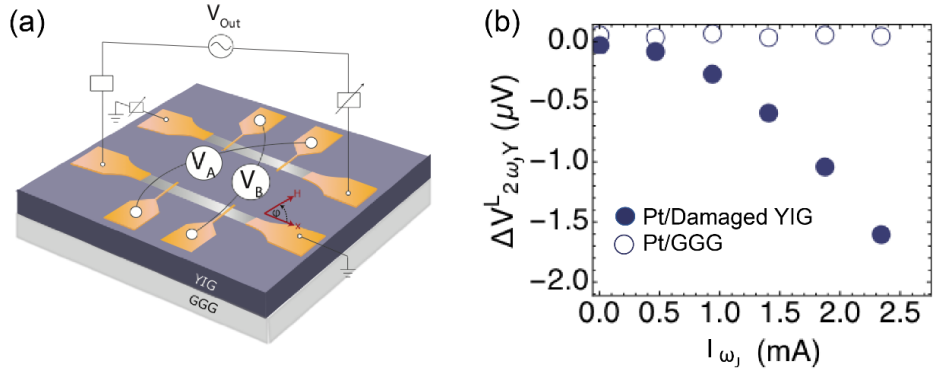

**FIG. S5**. Lock-in magnetoresistance measurements on a damaged YIG/GGG substrate and a GGG substrate. (a) Schematic illustration of longitudinal bridge measurements of two parallel Pt lines. (b) The measured $\Delta V_{2\omega_J,Y} = V_{A,2\omega_J,Y} - V_{B,2\omega_J,Y}$ as a function of the rms first-harmonic amplitude ($I_{\omega_J}$) of the applied charge current for the Pt/YIG sample (filled symbols) and a Pt/GGG control device (unfilled symbols) for $\varphi = -90°$.

### H. Frequency-dependent Longitudinal Wheatstone Bridge Measurement

A Zurich Instrument ultrahigh frequency lock-in (UHFLI) amplifier was used in the frequency-dependent longitudinal Wheatstone bridge measurement. As shown in Fig. S6(a), the constant voltage output ($V_L$) of the UHFLI with an output impedance ($R_L$) of 50 Ω was connected to the two middle Hall probes of the Pt line. The two end contacts to the Pt lines were connected with a 100 Ω precision resistor ($R_1$) and a variable resistor ($R_2$), respectively, to the ground. In this connection, the current flows in the two segments ($R_{S1}$ and $R_{S2}$) of the Pt line are antiparallel to each other, yielding opposite spin Peltier effects in the two segments. The voltage drops ($V_1$ and $V_2$) over $R_1$ and $R_2$ were connected to the inputs of a high-frequency, unitary-gain AD8130 differential-to-single ended converter, the output ($v$) of which was measured with a single-ended input channel of the UHFLI.

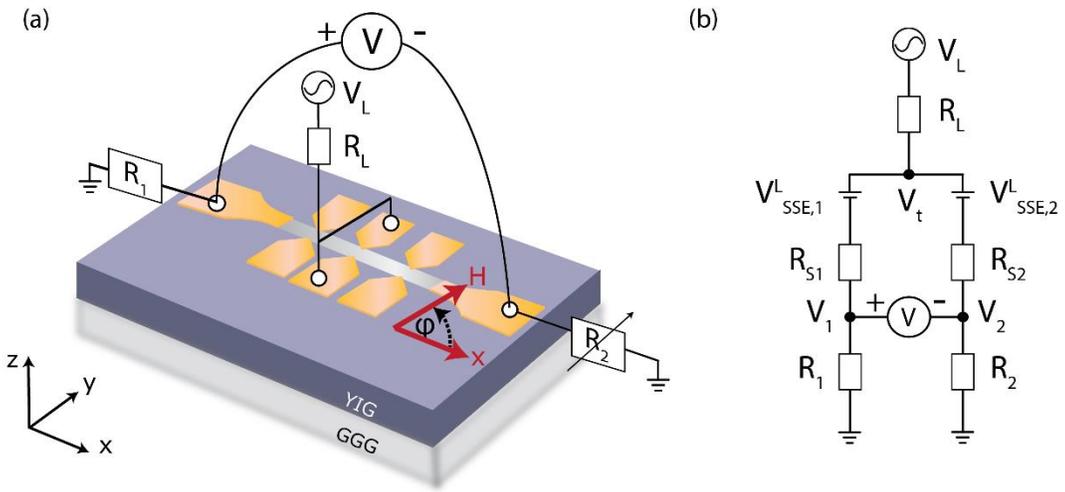

**FIG. S6**. Wheatstone balanced bridge measurements. (a) Schematic and (b) circuit diagram of the Wheatstone balanced bridge with a constant voltage source

Figure S6(b) shows the equivalent Wheatstone bridge circuit, where

$$\frac{V_L - V_t}{R_L} = \frac{V_t - V_{SSE,1}^L}{R_{S1} + R_1} + \frac{V_t - V_{SSE,2}^L}{R_{S2} + R_2} \quad (S42)$$

Thus,

$$V_t = \frac{V_L}{1 + \frac{R_L}{R_{S1} + R_1} + \frac{R_L}{R_{S2} + R_2}} + \frac{V_{SSE,1}^L}{1 + \frac{R_{S1} + R_1}{R_{S2} + R_2} + \frac{R_{S1} + R_1}{R_L}} + \frac{V_{SSE,2}^L}{1 + \frac{R_{S2} + R_2}{R_{S1} + R_1} + \frac{R_{S2} + R_2}{R_L}} \quad (S43)$$

The bridge voltage is obtained as

$$V = V_1 - V_2 = \frac{R_1}{R_{S1} + R_1}\left(V_t - V_{SSE,1}^L\right) - \frac{R_2}{R_{S2} + R_2}\left(V_t - V_{SSE,2}^L\right) \quad (S44)$$

which is expanded into the following bridge circuit equation

$$V[R_1R_2 + (R_1 + R_2)R_L + (R_2 + R_L)R_{S1} + (R_1 + R_L)R_{S2}+R_{S1}R_{S2}]$$
$$= V_L (R_1R_{S2} - R_2R_{S1})$$
$$- V_{SSE,1}^L[R_1R_{S2} + (R_1+R_2)R_L + R_1R_2]+V_{SSE,2}^L[R_2R_{S1} + (R_1 + R_2)R_L$$
$$+ R_1R_2] \qquad (S45)$$

The applied voltage is modulated according to

$$V_L = \sqrt{2}V_{L,\omega_J} \sin \omega_J t \qquad (S46)$$

The Pt line resistance contains a first harmonic unidirectional magnetoresistance (UMR) component, which is mainly caused by the SPE, in addition to the steady and second harmonic components caused by Joule heating,

$$R_{S1} = R_{S1,0} + \sqrt{2}\big(R_{S1,\omega_J,X} \sin \omega_J t + R_{S1,\omega_J,Y} \cos \omega_J t + R_{S1,2\omega_J,X} \sin 2\omega_J t$$
$$+ R_{S1,2\omega_J,Y} \cos 2\omega_J t\big) \qquad (S47)$$

and

$$R_{S2} = R_{S2,0} + \sqrt{2}\big(R_{S2,\omega_J,X} \sin \omega_J t + R_{S2,\omega_J,Y} \cos \omega_J t + R_{S2,2\omega_J,X} \sin 2\omega_J t$$
$$+ R_{S2,2\omega_J,Y} \cos 2\omega_J t\big) \qquad (S48)$$

Due to the opposite SPE temperature rise in the two Pt segments with the same temperature coefficient of resistance,

$$\frac{R_{S2,\omega_J,X}}{R_{S1,\omega_J,X}} = \frac{R_{S2,\omega_J,Y}}{R_{S1,\omega_J,Y}} = -\frac{R_{S2,4p}}{R_{S1,4p}} \equiv -\gamma \qquad (S49)$$

where $R_{S1,4p}$ and $R_{S2,4p}$ are the four-probe resistances of the two Pt line segments. In comparison, the temperature rise caused by Joule heating is the same for the two Pt line segments so that

$$\frac{R_{S2,2\omega_J,X}}{R_{S1,2\omega_J,X}} = \frac{R_{S2,2\omega_J,Y}}{R_{S1,2\omega_J,Y}} = \frac{R_{S2,4p}}{R_{S1,4p}} = \gamma \qquad (S50)$$

The SSE voltage $V_{SSE,1}^L$ contains a steady component ($V_{SSE,0}^L$) and a second harmonic component caused by Joule heating, as well as a negligibly small first-harmonic component caused by the SPE,

$$V_{SSE,1}^L = -L_h E_{SSE}^L(\varphi = -90°) = V_{SSE,0}^L + \sqrt{2}\left(V_{SSE,\omega_J,X}^L \sin \omega_J t + V_{SSE,\omega_J,Y}^L \cos \omega_J t + \right.$$
$$\left. V_{SSE,2\omega_J,X}^L \sin 2\omega_J t + V_{SSE,2\omega_J,Y}^L \cos 2\omega_J t\right) \qquad (S51)$$

where $L_h$ is the length of the half Pt line segment. Because of the opposite current flow directions in the two Pt line segments in the bridge measurement and the opposing polarities defined for the two SSE voltage in Fig. S6,

$$V_{SSE,2}^L = -V_{SSE,0}^L$$
$$+ \sqrt{2}\left(V_{SSE,\omega_J,X}^L \sin\omega_J t + V_{SSE,\omega_J,Y}^L \cos\omega_J t - V_{SSE,2\omega_J,X}^L \sin 2\omega_J t\right.$$
$$\left. - V_{SSE,2\omega_J,Y}^L \cos 2\omega_J t\right) \qquad (S52)$$

The bridge voltage contains different frequency components as

$$V = V_0 + \sqrt{2}\big(V_{\omega_J,X} \sin\omega_J t + V_{\omega_J,Y} \cos\omega_J t + V_{2\omega_J,X} \sin 2\omega_J t + V_{2\omega_J,Y} \cos 2\omega_J t$$
$$+ V_{3\omega_J,X} \sin 3\omega_J t + V_{3\omega_J,Y} \cos 3\omega_J t + V_{4\omega_J,X} \sin 4\omega_J t + V_{4\omega_J,Y} \cos 4\omega_J t\big) \quad (S53)$$

We define $\xi_+ \equiv \xi(+|\mathbf{H}|)$, $\xi_- \equiv \xi(-|\mathbf{H}|)$, $\xi_m \equiv 0.5(\xi_+ + \xi_-)$, $\Delta\xi \equiv \xi_+ - \xi_-$. The sign of the SPE depends on the direction of an in-plane magnetic field $\mathbf{H}$ that is perpendicular to the current. Thus, $R_{S1,\omega_J,X,+} = -R_{S1,\omega_J,X,-} \equiv R_{\omega_J,X}$ and $R_{S1,\omega_J,Y+} = -R_{S1,\omega_J,Y-} \equiv R_{\omega_J,Y}$. In comparison, $R_{S1,2\omega_J,X,+} = R_{S1,2\omega_J,X,-} \equiv R_{2\omega_J,X}$ and $R_{S1,2\omega_J,Y,+} = R_{S1,2\omega_J,Y,-} \equiv R_{2\omega_J,Y}$ because Joule heating is independent of the magnetic field. Similarly, $V_{SSE,\omega_J}^L(|H|) = V_{SSE,\omega_J}^L(-|H|)$, $V_{SSE,2\omega_J}^L(|H|) = -V_{SSE,2\omega_J}^L(-|H|)$, and $V_{SSE,0}^L(|H|) = V_{SSE,0}^L(-|H|)$.

Subtracting the bridge voltage equations at two opposite magnetic fields leads to two equations for the coefficients of the $\sin 2\omega_J t$ and $\cos 2\omega_J t$ components. The two equations can be used to obtain the exact solution of $R_{\omega_J,X}$ and $R_{\omega_J,Y}$. The exact solution differs within 0.3% from the following approximation that ignores the higher order terms

$$R_{\omega_J,X} \approx R_0^2 \frac{\Delta V_{2\omega_J,Y} - \gamma_L \Delta V_{SSE,2\omega_J,Y}^L}{\sqrt{2}(\gamma R_1 + R_2)V_{L,rms}} \qquad (S54)$$

and

$$R_{\omega_J,Y} \approx -R_0^2 \frac{\Delta V_{2\omega_J,X} - \gamma_L \Delta V_{SSE,2\omega_J,X}^L}{\sqrt{2}(\gamma R_1 + R_2)V_{L,rms}} \qquad (S55)$$

where $\Delta V_{SSE,2\omega_J}^L = -(L_h/w_H)\Delta V_{SSE,2\omega_J}^T$,

$$R_0^2 \equiv R_1 R_2 + (R_1 + R_2)R_L + (R_2 + R_L)R_{S1,0} + (R_1 + R_L)R_{S2,0} + R_{S1,0}R_{S2,0} \qquad (S56)$$

and

$$\gamma_L \equiv -\frac{2(R_1+R_2)R_L + 2R_1R_2 + R_2R_{S1,0} + R_1R_{S2,0}}{R_0^2} \tag{S57}$$

To obtain $R_{2\omega_J,X}$ and $R_{2\omega_J,Y}$, we carried out additional third harmonic bridge measurements by replacing the $R_{S2}$ segment of the Pt line with a precision resistor $R'_{S2}$. In addition, the magnetic field is parallel to the current flow in the Pt line to eliminate the SPE and SEE in this longitudinal measurement, so that $R_{S1}$ only contains the Joule heating effect that is independent of the magnetic field direction according to

$$R_{S1} = R_{S1,0} + \sqrt{2}\left(R_{2\omega_J,X} \sin 2\omega_J t + R_{2\omega_J,Y} \cos 2\omega_J t\right) \tag{S58}$$

The measured bridge voltage becomes

$$V' = V'_0 + \sqrt{2}\Big(V'_{\omega_J,X} \sin \omega_J t + V'_{\omega_J,Y} \cos \omega_J t + V'_{2\omega_J,X} \sin 2\omega_J t + V'_{2\omega_J,Y} \cos 2\omega_J t$$
$$+ V'_{3\omega_J,X} \sin 3\omega_J t + V'_{3\omega_J,Y} \cos 3\omega_J t + V'_{4\omega_J,X} \sin 4\omega_J t$$
$$+ V'_{4\omega_J,Y} \cos 4\omega_J t\Big) \tag{S59}$$

The bridge circuit equation is considerably simplified as

$$\Big[V_0 + \sqrt{2}\big(V'_{\omega_J,X} \sin \omega_J t + V'_{\omega_J,Y} \cos \omega_J t + V'_{2\omega_J,X} \sin 2\omega_J t + V'_{2\omega_J,Y} \cos 2\omega_J t$$
$$+ V'_{3\omega_J,X} \sin 3\omega_J t + V'_{3\omega_J,Y} \cos 3\omega_J t + V'_{4\omega_J,X} \sin 4\omega_J t$$
$$+ V'_{4\omega_J,Y} \cos 4\omega_J t\big)\Big]\Big[R'^2_0$$
$$+ \sqrt{2}(R_2 + R_L + R'_{S2})\big(R_{2\omega_J,X} \sin 2\omega_J t + R_{2\omega_J,Y} \cos 2\omega_J t\big)\Big]$$
$$= \sqrt{2}V_{L,rms} \sin \omega_J t \Big[R_1 R'_{S2} - R_2 R_{S1,0}$$
$$- \sqrt{2}R_2\big(R_{2\omega_J,X} \sin 2\omega_J t + R_{2\omega_J,Y} \cos 2\omega_J t\big)\Big] \tag{S60}$$

where

$$R'^2_0 \equiv R_1 R_2 + (R_1 + R_2)R_L + (R_2 + R_L)R_{S1,0} + (R_1 + R_L)R'_{S2} + R_{S1,0}R'_{S2} \tag{S61}$$

The $\sin 3\omega_J t$ and $\cos 3\omega_J t$ components, respectively, of this bridge circuit equation are identified as

$$(R_2 + R_L + R'_{S2})V'_{\omega_J,Y}R_{2\omega_J,X} + \big[R_2 V_{L,rms} + (R_2 + R_L + R'_{S2})V'_{\omega_J,X}\big]R_{2\omega_J,Y} = -\sqrt{2}R'^2_0 V'_{3\omega_J,X} \tag{S62}$$

and

$$\left[R_2 V_{L,rms}+(R_2+R_L+R'_{S2})V'_{\omega_J,X}\right]R_{2\omega_J,X}+(R_2+R_L+R'_{S2})V'_{\omega_J,Y}R_{2\omega_J,Y}=-\sqrt{2}R'^2_0 V'_{3\omega_J,Y}$$
(S63)

The solution is

$$R_{2\omega_J,X}=\sqrt{2}R'^2_0\frac{\left[R_2 V_{L,rms}+(R_2+R_L+R'_{S2})V'_{\omega_J,X}\right]V'_{3\omega_J,Y}-(R_2+R_L+R'_{S2})V'_{\omega_J,Y}V'_{3\omega_J,X}}{\left[R_2 V_{L,rms}+(R_2+R_L+R'_{S2})V'_{\omega_J,X}\right]^2+\left[(R_2+R_L+R'_{S2})V'_{\omega_J,Y}\right]^2}\approx\frac{\sqrt{2}R'^2_0 V'_{3\omega_J,Y}}{R_2 V_{L,rms}} \quad (S64)$$

and

$$R_{2\omega_J,Y}=-\sqrt{2}R'^2_0\frac{\left[R_2 V_{L,rms}+(R_2+R_L+R'_{S2})V'_{\omega_J,X}\right]V'_{3\omega_J,X}+(R_2+R_L+R'_{S2})V'_{\omega_J,Y}V'_{3\omega_J,Y}}{\left[R_2 V_{L,rms}+(R_2+R_L+R'_{S2})V'_{\omega_J,X}\right]^2+\left[(R_2+R_L+R'_{S2})V'_{\omega_J,Y}\right]^2}\approx-\frac{\sqrt{2}R'^2_0 V'_{3\omega_J,X}}{R_2 V_{L,rms}}$$
(S65)

The results are shown in Fig. S7.

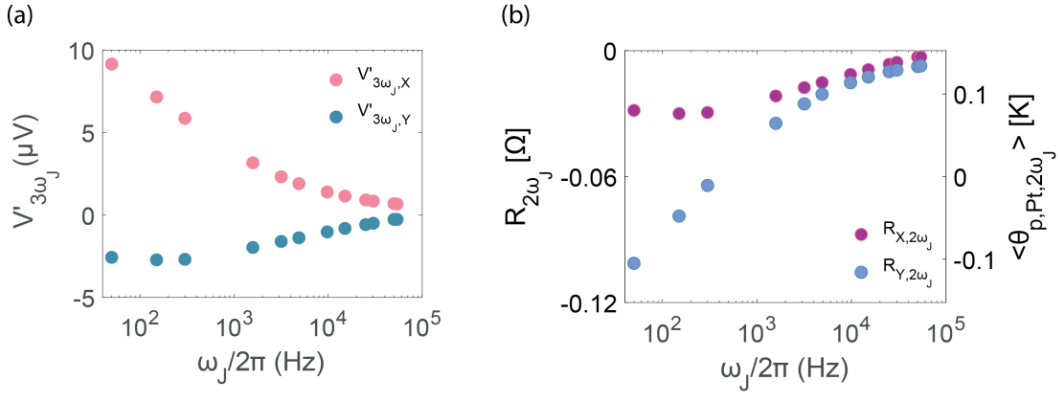

**FIG. S7**. Third harmonic ($3\omega_J$) bridge measurements. (a) The in-phase (X) and out-of-phase (Y) components of the measured third harmonic voltage ($V'_{3\omega_J}$) as a function of the frequency. (b) Second-harmonic modulation of the resistance (left axis) and equivalent phonon temperature (right axis) of the Pt line due to the Joule heating as a function of the current modulation frequency in the Pt.

We define

$$UMR_X\equiv\frac{R_{\omega_J,X}}{\frac{L_h}{L}R_{S1,4p}} \quad (S66)$$

and

$$UMR_Y \equiv \frac{R_{\omega_J,Y}}{\frac{L_h}{L}R_{S1,4p}} \tag{S67}$$

where $L$ is the distance between the two voltage probes used to measure the four-probe resistance, $R_{S1,4p}$, of the half Pt line with the two-probe resistance being $R_{S1}$. We calculate the corresponding rms amplitude of the Pt phonon temperature modulations caused by the SPE effect and Joule heating at the first and second harmonics, respectively, as

$$\langle \vartheta_{p,Pt,\omega_J,X} \rangle = \frac{UMR_X}{\text{TCR}} \tag{S68}$$

$$\langle \vartheta_{p,Pt,\omega_J,Y} \rangle = \frac{UMR_Y}{\text{TCR}} \tag{S69}$$

$$\langle \vartheta_{p,Pt,2\omega_J,X} \rangle = \frac{R_{2\omega_J,X}}{\frac{L_h}{L}R_{S1,4p}\text{TCR}} \tag{S70}$$

$$\langle \vartheta_{p,Pt,2\omega_J,Y} \rangle = \frac{R_{2\omega_J,Y}}{\frac{L_h}{L}R_{S1,4p}\text{TCR}} \tag{S71}$$

## I. Coupled Electron-Phonon Transport in Pt

In the following, $\kappa_p$ and $\kappa_e$ are the phonon and electron contributions, respectively, to the total thermal conductivity ($\kappa$), $c_p$ and $c_e$ are the phonon and electron contributions, respectively, to the total specific heat ($c$), and the Pt subscript is used to distinguish the phonon and total properties of Pt from the corresponding terms defined for YIG without this subscript. With the Thompson effect ignored inside the Pt, a unitary Fourier transform, $\hat{f}(\omega) = \frac{1}{\sqrt{2\pi}}\int_{-\infty}^{\infty}f(t)e^{-i\omega t}dt$, is applied to the two-temperature heat diffusion equations to obtain the angular frequency ($\omega$) components as

$$\nabla \cdot (\kappa_{p,pt}\nabla\hat{\vartheta}_{p,Pt}) + g_{ep}(\hat{\vartheta}_e - \hat{\vartheta}_{p,Pt}) = i\omega c_{p,Pt}\hat{\vartheta}_{p,Pt} \tag{S72}$$

and

$$\nabla \cdot (\kappa_e\nabla\hat{\vartheta}_{e,Pt}) + g_{ep}(\hat{\vartheta}_{p,Pt} - \hat{\vartheta}_e) + \hat{q} = i\omega c_e\hat{\vartheta}_e \tag{S73}$$

where $g_{ep}$ is the electron–phonon coupling constant, $q$ is the heat source term, $i = \sqrt{-1}$, $\vartheta \equiv T - T_\infty$, with $T$ and $T_\infty$ being the temperature and the ambient temperature, respectively. The electron-phonon thermalization length is

$$l_{ep} \equiv \left[g_{ep}\left(\frac{1}{\kappa_{p,Pt}} + \frac{1}{\kappa_e}\right)\right]^{-\frac{1}{2}} \tag{S74}$$

which is calculated as 2.8 nm. We define the following parameter with the same unit as the Pt thermal diffusivity ($\alpha_{Pt} \equiv \kappa_{Pt}/c_{Pt}$),

$$\alpha_{p,Pt} \equiv \frac{\kappa_{p,Pt}}{c_{p,Pt}} \tag{S75}$$

$$\alpha_e \equiv \frac{\kappa_e}{c_{e,Pt}} \tag{S76}$$

The thermal penetration depth in Pt is

$$\Lambda_{Pt} \equiv \left(\frac{\alpha_{Pt}}{\omega}\right)^{1/2} \tag{S77}$$

which exceeds 1 μm for $\frac{\omega}{2\pi} \leq 1 \times 10^5$ Hz. With these parameters

$$l_{ep}^2 \nabla^4 \hat{\vartheta}_{p,Pt} - \left[1 + i\left(\frac{\alpha_{Pt}}{\alpha_e} + \frac{\alpha_{Pt}}{\alpha_{p,Pt}}\right)\left(\frac{l_{ep}}{\Lambda_{Pt}}\right)^2\right] \nabla^2 \hat{\vartheta}_{p,Pt} + \frac{i}{\Lambda_{Pt}^2}\left[1 + \frac{i\alpha_{Pt}^2}{\alpha_{p,Pt}\alpha_e}\left(\frac{l_{ep}}{\Lambda_{Pt}}\right)^2\right] \hat{\vartheta}_{p,Pt} - \frac{\hat{q}_\omega}{\kappa_{Pt}} = 0 \tag{S78}$$

For 1D transport along the $z$ direction, the solution for the phonon temperature distributions in the Pt thin film is

$$\hat{\vartheta}_{p,Pt} = \sum_{j=0,1}\left[a_{j,Pt}\sinh\left(\frac{z}{l_{j,Pt}}\right) + b_{j,Pt}\cosh\left(\frac{z}{l_{j,Pt}}\right)\right] - \frac{i\Lambda_{Pt}^2 \hat{q}_\omega l}{\kappa_{Pt} l\left[1 + \frac{i\alpha_{Pt}^2}{\alpha_{p,Pt}\alpha_{e,Pt}}\left(\frac{l_{ep}}{\Lambda_{Pt}}\right)^2\right]} \tag{S79}$$

where the coefficients $a_{j=0,1,Pt}$ and $b_{j=0,1,Pt}$, are determined from the boundary conditions and

$$\frac{1}{l_{j=0,1,Pt}} = \frac{1}{l_{ep}}\sqrt{\frac{1}{2} + \frac{i}{2}\left(\frac{l_{ep}}{\Lambda_{Pt}}\right)^2\left(\frac{\alpha_{Pt}}{\alpha_e} + \frac{\alpha_{Pt}}{\alpha_{p,Pt}}\right) \pm \frac{(-1)^j}{2}\sqrt{1 + 2i\left(\frac{l_{ep}}{\Lambda_{Pt}}\right)^2\left(\frac{\alpha_{Pt}}{\alpha_e} + \frac{\alpha_{Pt}}{\alpha_{p,Pt}} - 2\right) - \left(\frac{l_{ep}}{\Lambda_{Pt}}\right)^4\left(\frac{\alpha_{Pt}}{\alpha_{p,Pt}} - \frac{\alpha_{Pt}}{\alpha_e}\right)^2}} \tag{S80}$$

In addition,

$$\hat{\vartheta}_{e,Pt} = \sum_{j=0,1} \zeta_{j,Pt}\left[a_{j,Pt}\sinh\left(\frac{z}{l_{j,Pt}}\right) + b_{j,Pt}\cosh\left(\frac{z}{l_{j,Pt}}\right)\right] - \left[1 + \frac{i\alpha_{Pt}\kappa_{Pt}}{\alpha_{p,Pt}\kappa_{e,Pt}}\left(\frac{l_{ep}}{\Lambda_{Pt}}\right)^2\right]\frac{i\Lambda_{Pt}^2 \hat{q} l}{\kappa_{Pt} l\left[1 + \frac{i\alpha_{Pt}^2}{\alpha_{p,Pt}\alpha_{e,Pt}}\left(\frac{l_{ep}}{\Lambda_{Pt}}\right)^2\right]} \tag{S81}$$

where

$$\zeta_{j,Pt} = \left[1 - \frac{\kappa_{pt}}{\kappa_{e,Pt}}\left(\frac{l_{ep}}{l_{j,Pt}}\right)^2 + \frac{i\alpha_{Pt}\kappa_{Pt}}{\alpha_{p,Pt}\kappa_{e,Pt}}\left(\frac{l_{ep}}{\Lambda_{Pt}}\right)^2\right] \quad (S82)$$

**J. Magnon Boltzmann Transport Equation**

The magnon occupation ($f_m$) of wavevector $\boldsymbol{k}$ at location $\boldsymbol{r}$ and time $t$ is governed by the Boltzmann transport equation (BTE),

$$\frac{\partial f_m}{\partial t} + \boldsymbol{v}\cdot\nabla f_m = \left(\frac{\partial f_m}{\partial t}\right)_s \quad (S83)$$

where $\boldsymbol{v}$ is the group velocity of the magnon mode, and the term in the right hand side is caused by different magnon scattering processes. Elastic scattering processes that converse the magnon number density ($n_m$) and energy density ($u$) relax $f_m(\boldsymbol{r},\boldsymbol{k})$ towards the following quasi-equilibrium distribution that are parametrized by the local magnon temperature $T_m$ and chemical potential $\mu_m$

$$\overline{f_m}(\omega) = \frac{1}{\exp\left[\frac{\hbar(\omega(\boldsymbol{k}) + \Omega) - \mu_m(\boldsymbol{r})}{k_B T_m(\boldsymbol{r})}\right] - 1} \quad (S84)$$

which satisfies

$$n_m = \frac{1}{V}\sum_{\boldsymbol{k}} f_m(\boldsymbol{k}) = \frac{1}{V}\sum_{\boldsymbol{k}} \overline{f_m}(\omega) \quad (S85)$$

and

$$u = \frac{1}{V}\sum_{\boldsymbol{k}} f_m(\boldsymbol{r},\boldsymbol{k})\hbar(\omega(\boldsymbol{k}) + \Omega) = \frac{1}{V}\sum_{\boldsymbol{k}} \overline{f_m}(\omega)\hbar(\omega(\boldsymbol{k}) + \Omega) \quad (S86)$$

Here, $k_B$ is the Boltzmann constant, $\omega(\boldsymbol{k})$ is the angular frequency of the magnon mode $\boldsymbol{k}$, $\hbar\Omega$ is the magnon gap that depends on the external applied magnetic field ($H$), $V$ is the volume, and the summation is over all modes in the first Brillouin zone. The chemical potential arises because $n_m$ is conserved within the time scale ($\tau_c$) for such elastic scattering processes.

Within a longer time scale ($\tau_{mp}$) than $\tau_c$, inelastic spin-conserving magnon-phonon scattering processes equilibrate $T_m$ with the local phonon bath temperature ($T_p$) and relax $f_m(\boldsymbol{r},\boldsymbol{k})$ towards $\overline{f_{m,0}}(\omega) = \overline{f_m}(\omega, T_m = T_p)$. At an even longer time scale ($\tau_\mu$) than $\tau_{mp}$, quasielastic scattering processes that do not conserve $n$ nullify the magnon chemical potential and relax $f_m(\boldsymbol{r},\boldsymbol{k})$ towards $f_{m,0}(\omega) = \overline{f_m}(\omega, \mu_m = 0)$. These different scattering rates can be summed up with the relaxation time approximation to obtain

$$\left(\frac{\partial f_m(\boldsymbol{r},\boldsymbol{k})}{\partial t}\right)_s \approx \frac{\overline{f_m}(\omega) - f_m(\boldsymbol{r},\boldsymbol{k})}{\tau_c(\boldsymbol{k})} + \frac{\overline{f_{m,0}}(\omega) - f_m(\boldsymbol{r},\boldsymbol{k})}{\tau_{mp}(\boldsymbol{k})} + \frac{f_{m,0}(\omega) - f_m(\boldsymbol{r},\boldsymbol{k})}{\tau_\mu(\boldsymbol{k})} \qquad (S87)$$

With the assumption of $\nabla f_m(\boldsymbol{r},\boldsymbol{k}) \approx \nabla \overline{f_m}(\omega)$ and $\frac{\partial f_m(\boldsymbol{r},\boldsymbol{k})}{\partial t} \approx \frac{\partial \overline{f_m}(\omega)}{\partial t}$ and under a constant magnetic field,

$$f_m(\boldsymbol{r},\boldsymbol{k}) = \frac{\left\{\frac{\overline{f_m}(\omega)}{\tau_c} + \frac{\overline{f_{m,0}}(\omega)}{\tau_{mp}} + \frac{f_{m,0}(\omega)}{\tau_\mu} - \boldsymbol{v}(\boldsymbol{k}) \cdot \left[\frac{\partial \overline{f_m}(\omega)}{\partial \mu_m}\nabla\mu_m + \frac{\partial \overline{f_m}(\omega)}{\partial T_m}\nabla T_m\right] - \frac{\partial \overline{f_m}(\omega)}{\partial t}\right\}}{\left(\tau_c^{-1} + \tau_{mp}^{-1} + \tau_\mu^{-1}\right)} \qquad (S88)$$

**K. Magnon Spin Diffusion**

The magnon particle flux can be calculated as

$$J_n = \frac{1}{V}\sum_{\boldsymbol{k}} \boldsymbol{v}(\boldsymbol{k}) f_m(\boldsymbol{r},\boldsymbol{k}) = -\sigma_m(\nabla\mu_m + S_m \nabla T_m) \qquad (S89)$$

where the summations over $f_{m,0}(\omega)$, $\overline{f_m}(\omega)$ and $\overline{f_{m,0}}(\omega)$ vanish to yield the magnon conductivity ($\sigma_m$) and bulk spin Seebeck coefficient ($S_m$) as

$$\sigma_m = \frac{1}{V}\sum_{\boldsymbol{k}} v_\parallel^2 \frac{\partial \overline{f_m}(\omega)}{\partial \mu_m}\left(\tau_c^{-1} + \tau_{mp}^{-1} + \tau_\mu^{-1}\right)^{-1} \qquad (S90)$$

$$S_m = \frac{1}{\sigma_m V}\sum_{\boldsymbol{k}} v_\parallel^2 \frac{\partial \overline{f_m}(\omega)}{\partial T_m}\left(\tau_c^{-1} + \tau_{mp}^{-1} + \tau_\mu^{-1}\right)^{-1} \qquad (S91)$$

Here, $v_\parallel$ is the velocity component that is parallel to the corresponding gradient that acts as the driving force.

The rate of change of the magnon density can be calculated as

$$\frac{\partial n_m}{\partial t} = \frac{1}{V}\frac{\partial \sum_{\boldsymbol{k}} f_m(\boldsymbol{r},\boldsymbol{k})}{\partial t} = -\frac{1}{V}\sum_{\boldsymbol{k}}(\boldsymbol{v}(\boldsymbol{k}) \cdot \nabla f_m(\boldsymbol{r},\boldsymbol{k})) + \frac{1}{V}\left(\frac{\partial \sum_{\boldsymbol{k}} f_m(\boldsymbol{r},\boldsymbol{k})}{\partial t}\right)_s \qquad (S92)$$

The magnon group velocity for mode $\boldsymbol{k}$ is assumed to be spatially uniform,

$$\nabla \boldsymbol{v}(\boldsymbol{k}) = 0 \qquad (S93)$$

which is used to obtain the magnon spin diffusion equation as

$$\frac{\partial n_m}{\partial t} + \nabla \cdot J_n = \left(\frac{\partial n_m}{\partial t}\right)_s \qquad (S94)$$

Under the relaxation time approximation,

$$\left(\frac{\partial n_m}{\partial t}\right)_S = \frac{1}{V}\left(\frac{\partial \sum_k f_m(\bm{r},\bm{k})}{\partial t}\right)_S \approx -\frac{1}{\widetilde{\tau_{n,\mu}}}\frac{\partial n_m}{\partial \mu_m}\mu_m \approx \frac{n_{m,0} - n_m}{\widetilde{\tau_n}} \qquad (S95)$$

Here, $n_m$ relaxes to $n_{m,0}$ at a time scale $\widetilde{\tau_n}$ close to the average ($\widetilde{\tau_{n,\mu}}$) of $\tau_\mu(\bm{k})$, because the magnon temperature is relaxed by magnon-phonon scattering processes that mostly conserve the number of magnons. In addition,

$$n_{m,0} = \frac{1}{V}\sum_k f_{m,0}(\bm{r},\bm{k}) \qquad (S96)$$

Thus, the magnon spin diffusion equation can be expressed as a function of the driving forces as

$$\frac{\partial n_m}{\partial \mu_m}\frac{\partial\left(\mu_m - \hbar\frac{\partial\Omega}{\partial H}H\right)}{\partial t} + \frac{\partial n_m}{\partial T_m}\frac{\partial T_m}{\partial t} - \nabla\cdot\left[\sigma_m\left(\nabla\mu_m - \hbar\frac{\partial\Omega}{\partial H}\nabla H + S_m\nabla T_m\right)\right] \approx -\frac{1}{\widetilde{\tau_{n,\mu}}}\frac{\partial n_m}{\partial \mu_m}\mu_m \quad (S97)$$

Here,

$$\frac{\partial n_m}{\partial T_m} = -\frac{1}{g\mu_B}\frac{\partial M}{\partial T_m} \qquad (S98)$$

where $M$ is the magnetization, $g$ is the Landé $g$ factor, and $\mu_B$ is the Bohr magneton. In addition, an analytical approximation of the magnon dispersion can be used to calculate [3,12]

$$\frac{\partial n_m}{\partial \mu_m} = \sqrt{k_B T}Li_{\frac{1}{2}}\left(e^{-\left(\frac{\hbar\Omega}{k_B T}\right)}\right)\Lambda^3 \qquad (S99)$$

where $Li_{\frac{1}{2}}$ is the polylogarithmic function of order 1/2 and the value of the magnon gap energy is $\hbar\Omega = 4.8$ μeV [12,13]. Under the gray approximation, the velocity and relaxation time terms are replaced by the corresponding average values ($\overline{v_\|^2}, \overline{\tau_c}, \overline{\tau_{mp}}$, and $\overline{\tau_\mu}$) and taken out of the summation to obtain

$$\bm{J}_n \approx -\overline{v_\|^2}\cdot\left(\overline{\tau_c}^{-1} + \overline{\tau_{mp}}^{-1} + \overline{\tau_\mu}^{-1}\right)^{-1}\frac{1}{V}\sum_k\left(\frac{\partial \overline{f_m}(\omega)}{\partial\mu_m}\nabla\mu_m + \frac{\partial\overline{f_m}(\omega)}{\partial T_m}\nabla T_m\right)$$

$$= -D\left(\frac{\partial n_m}{\partial\mu_m}\nabla\mu_m + \frac{\partial n_m}{\partial T_m}\nabla T_m\right) \approx -D_n\nabla n_m \qquad (S100)$$

where $D_n = \overline{v_\|^2}\left(\overline{\tau_c}^{-1} + \overline{\tau_{mp}}^{-1} + \overline{\tau_\mu}^{-1}\right)^{-1}$ is equivalent to the average diffusivity of magnons,

$$\sigma_m \approx D_n\frac{\partial n_m}{\partial\mu_m} \qquad (S101)$$

In addition, the magnon spin diffusion length is defined as $\lambda \equiv (D_n\widetilde{\tau_n})^{1/2}$.

### L. Magnon Heat Diffusion

The magnon energy flux can be calculated as

$$J_{u,m} = \frac{1}{V}\sum_{k} v(k) f_m(r,k) \hbar(\omega(k) + \Omega) = J_{Q,m} + \mu_m J_n \tag{S102}$$

where the magnon heat flux

$$J_{Q,m} = \frac{1}{V}\sum_{k} v(k) f_m(r,k) [\hbar(\omega(k) + \Omega) - \mu_m(r)] = -\sigma_m \Gamma_m \nabla \mu_m - \kappa_m \nabla T_m$$

$$= \Gamma_m J_n - \kappa_M \nabla T_m \tag{S103}$$

Based on $\frac{\partial \overline{f_m}(\omega)}{\partial T_m} = \frac{\hbar(\omega(k)+\Omega)-\mu(r)}{T_m} \frac{\partial \overline{f_m}(\omega)}{\partial \mu_m}$, we obtain the Onsager reciprocal relation for the magnon bulk spin Peltier coefficient ($\Gamma_m$) as

$$\Gamma_m = \frac{1}{\sigma_m V}\sum_{k} v(k)_\parallel^2 \frac{\partial \overline{f_m}(\omega)}{\partial \mu_m} [\hbar(\omega(k) + \Omega) - \mu_m(r)] \left(\tau_c^{-1} + \tau_{mp}^{-1} + \tau_\mu^{-1}\right)^{-1} = T_m S_m \tag{S104}$$

In addition, the magnon thermal conductivity is defined for both the close-circuit ($\kappa_m$) and open-circuit ($\kappa_M$) conditions as

$$\kappa_m = \frac{1}{V}\sum_{k} v(k)_\parallel^2 \frac{\partial \overline{f_m}(\omega)}{\partial T_m} [\hbar(\omega(k) + \Omega) - \mu_m(r)] \left(\tau_c^{-1} + \tau_{mp}^{-1} + \tau_\mu^{-1}\right)^{-1} \tag{S105}$$

and

$$\kappa_M = \kappa_m - \sigma_m T_m S_m^2 \geq 0 \tag{S106}$$

which sets an upper limit for $S_m$. The rate of change of the magnon energy density is calculated as

$$\frac{\partial u}{\partial t} = \frac{1}{V} \frac{\partial \sum_k f_m(r,k)[\hbar(\omega(k) + \Omega)]}{\partial t}$$

$$= -\frac{1}{V}\sum_{k}[(v(k) \cdot \nabla f_m(r,k))][\hbar(\omega(k) + \Omega)] + \frac{1}{V}\left(\frac{\partial \sum_k [\hbar(\omega(k) + \Omega)] f_m(r,k)}{\partial t}\right)_S \tag{S107}$$

Based on the relaxation time approximation, the magnons exchange their energy with phonons at a rate of

$$\frac{1}{V}\left(\frac{\partial \sum_k [\hbar(\omega(k) + \Omega)] f_m(r,k)}{\partial t}\right)_S \approx -\frac{c_m}{\widetilde{\tau_{u,mp}}} (T_m - T_p) \tag{S108}$$

Here, the magnon chemical potential is relaxed via quasielastic scattering processes that do not conserve the number of magnons and but mostly conserve magnon energy, and $\widetilde{\tau_{u,mp}}$ is the weighted average of $\tau_{mp}(k)$. In addition, the magnon specific heat is

$$c_m = \frac{\partial u}{\partial T_m} \tag{S109}$$

The magnon-phonon coupling constant is $g_{mp} \equiv \frac{C_m}{\widetilde{\tau_{u,mp}}}$, and

$$\frac{\partial u}{\partial \mu_m} = \frac{1}{V} \sum_k \frac{\partial \overline{f_m}(\omega)}{\partial \mu_m} \hbar(\omega(\mathbf{k}) + \Omega) = \frac{1}{V} \sum_k \frac{\partial \overline{f_m}(\omega)}{\partial \mu_m} [\hbar(\omega(\mathbf{k}) + \Omega) - \mu_m] + \frac{\mu_m}{V} \sum_k \frac{\partial \overline{f_m}(\omega)}{\partial \mu_m}$$

$$= T_m \frac{\partial n_m}{\partial T_m} + \mu_m \frac{\partial n_m}{\partial \mu_m} \tag{S110}$$

where the second term can be ignored as $\frac{\partial u}{\partial \mu_m}$ is evaluated here for $\mu_m$ relaxation toward 0.

Meanwhile,

$$\frac{1}{V} \sum_k [(\mathbf{v}(\mathbf{k}) \cdot \nabla f_m(\mathbf{r}, \mathbf{k}))][\hbar(\omega(\mathbf{k}) + \Omega)] = \nabla \cdot \mathbf{J}_{u,m} = -\nabla \cdot (\kappa_M \nabla T_m) + \nabla \cdot [(\Gamma_m + \mu_m)\mathbf{J}_n] \tag{S111}$$

Here,

$$\nabla \cdot [(\Gamma_m + \mu_m)\mathbf{J}_n] = (\Gamma_m + \mu_m)\nabla \cdot \mathbf{J}_n + \mathbf{J}_n \cdot (\nabla \Gamma_m + \nabla \mu_m)$$

$$= (\Gamma_m + \mu_m)\nabla \cdot \mathbf{J}_n + \mathbf{J}_n \cdot \left(\nabla \Gamma_m - \frac{\mathbf{J}_n}{\sigma_m} - S_m \nabla T_m\right)$$

$$= (\Gamma_m + \mu_m)\nabla \cdot \mathbf{J}_n + \mathbf{J}_n \cdot \left(T_m \nabla S_m - \frac{\mathbf{J}_n}{\sigma_m}\right) \tag{S112}$$

which is used to express the magnon energy equation as

$$\frac{\partial u}{\partial \mu_m} \frac{\partial (\mu_m)}{\partial t} + c_m \frac{\partial T_m}{\partial t}$$

$$\approx \nabla \cdot (\kappa_M \nabla T_m) - (T_m S_m + \mu_m)\nabla \cdot \mathbf{J}_n + \frac{\mathbf{J}_n \cdot \mathbf{J}_n}{\sigma_m} - \mathbf{J}_n \cdot (T_m \nabla S_m) - \frac{C_m}{\widetilde{\tau_{u,mp}}}(T_m - T_p) \tag{S113}$$

With the $\nabla S_m$ term being equivalent to the Thompson effect, this equation takes a similar form as the electron energy equation, except that $\nabla \cdot \mathbf{J}_n$ does not vanish and is instead given by the magnon spin diffusion equation as

$$\nabla \cdot \mathbf{J}_n \approx -\frac{\partial n}{\partial T_m} \frac{\partial T_m}{\partial t} - \frac{\partial n}{\partial \mu_m}\left(\frac{\mu_m}{\widetilde{\tau_{n,\mu}}} + \frac{\partial \mu_m}{\partial t}\right) \tag{S114}$$

The $\mathbf{J}_n$ term is expressed as a function of the driving forces to obtain the magnon heat diffusion equation as

$$\frac{\partial u}{\partial \mu_m} \frac{\partial (\mu_m)}{\partial t} + c_m \frac{\partial T_m}{\partial t}$$

$$\approx \nabla \cdot [(\kappa_m + \mu_m \sigma_m S_m)\nabla T_m] + \nabla \cdot [(T_m S_m + \mu_m)\sigma_m \nabla \mu_m] - \frac{C_m}{\widetilde{\tau_{u,mp}}}(T_m - T_p) \tag{S115}$$

## M. Coupled Magnon Spin and Heat Diffusion Equations

When $H$, $\sigma_m$, and $\sigma_m S_m$ are assumed to be constant, the magnon spin diffusion equation becomes

$$\sigma_m \nabla^2 \mu_m + \sigma_m S_m \nabla^2 T_m \approx \frac{\partial n_m}{\partial \mu_m}\left(\frac{\partial \mu_m}{\partial t} + \frac{\mu_m}{\widetilde{\tau_{n,\mu}}}\right) + \frac{\partial n_m}{\partial T_m}\frac{\partial T_m}{\partial t} \tag{S116}$$

When $H$, $\kappa_{m,\mu} \equiv \kappa_m + \mu_m \sigma_m S_m$, and $\Gamma_{m,\mu}\sigma_m \equiv (\Gamma_m + \mu_m)\sigma_m$ are assumed to be constant, the magnon heat diffusion equation is reduced to

$$\Gamma_{m,\mu}\sigma_m \nabla^2 \mu_m + \kappa_{m,\mu}\nabla^2 T_m \approx \frac{\partial u}{\partial \mu_m}\frac{\partial \mu_m}{\partial t} + c_m\left(\frac{\partial T_m}{\partial t} + \frac{T_m - T_p}{\widetilde{\tau_{u,mp}}}\right) \tag{S117}$$

where $\Gamma_{m,\mu} \approx \Gamma_m$ and $\kappa_{m,\mu} \approx \kappa_m$ because $\mu_m \ll \Gamma_m$. These two equations are similar to the matrix expression in a prior report [3]. The magnon-phonon energy exchange terms appear in the following phonon heat diffusion equation

$$\kappa_p \nabla^2 T_p \approx -c_m \frac{T_m - T_p}{\widetilde{\tau_{u,mp}}} + c_p \frac{\partial T_p}{\partial t} \tag{S118}$$

With the use of $\vartheta \equiv T - T_\infty$, unitary Fourier transform is operated on the three equations to obtain

$$\sigma_m \nabla^2 \hat{\mu}_m + \sigma_m S_m \nabla^2 \hat{\vartheta}_m \approx \frac{1}{\widetilde{\tau_{n,\mu}}}\left(1 + i\omega\widetilde{\tau_{n,\mu}}\right)\frac{\partial n_m}{\partial \mu_m}\hat{\mu}_m + i\omega\hat{\vartheta}_m\frac{\partial n}{\partial T_m} \tag{S119}$$

$$\Gamma_{m,\mu}\sigma_m \nabla^2 \hat{\mu}_m + \kappa_{m,\mu}\nabla^2 \hat{\vartheta}_m \approx \frac{c_m}{\widetilde{\tau_{u,mp}}}\left(\hat{\vartheta}_m - \hat{\vartheta}_p\right) + i\omega\left(c_m \hat{\vartheta}_m + \frac{\partial u}{\partial \mu_m}\hat{\mu}_m\right) \tag{S120}$$

$$\kappa_p \nabla^2 \hat{\vartheta}_p \approx -\frac{c_m}{\widetilde{\tau_{u,mp}}}\left(\hat{\vartheta}_m - \hat{\vartheta}_p\right) + i\omega c_p \hat{\vartheta}_p \tag{S121}$$

The solution to the three coupled equations for the YIG film is

$$\hat{\vartheta}_p = \sum_{j=0,1,2}\left[a_j \exp\left(\frac{z}{l_j}\right) + b_j \exp\left(\frac{-z}{l_j}\right)\right] \tag{S122}$$

$$\hat{\vartheta}_{m-p} \equiv \hat{\vartheta}_m - \hat{\vartheta}_p = \sum_{j=0,1,2} e_j \left[a_j \exp\left(\frac{z}{l_j}\right) + b_j \exp\left(\frac{-z}{l_j}\right)\right] \tag{S123}$$

$$\hat{\mu}_m = \sum_{j=0,1,2} d_j \left[a_j \exp\left(\frac{z}{l_j}\right) + b_j \exp\left(\frac{-z}{l_j}\right)\right] \tag{S124}$$

Here, $a_j$ and $b_j$ are determined by the boundary conditions while the coefficients $l_j$, $e_j$, and $d_j$ satisfy the following requirements

$$\frac{\sigma_m d_j}{l_j^2} + \frac{\sigma_m S_m (e_j + 1)}{l_j^2} \approx \left(\frac{1}{\widetilde{\tau_{n,\mu}}} + i\omega\right) \frac{\partial n_m}{\partial \mu_m} d_j + i\omega \frac{\partial n_m}{\partial T_m} (e_j + 1) \tag{S125}$$

$$\frac{\Gamma_{m,\mu} \sigma_m d_j}{l_j^2} + \frac{\kappa_{m,\mu}(e_j + 1)}{l_j^2} \approx i\omega \frac{\partial u}{\partial \mu_m} d_j + \left(\frac{c_m}{\widetilde{\tau_{u,mp}}} + i\omega c_m\right) e_j + i\omega c_m \tag{S126}$$

$$\frac{\kappa_p}{l_{j,YIG}^2} \approx -\frac{c_m}{\widetilde{\tau_{u,mp}}} e_j + i\omega c_p \tag{S127}$$

Therefore,

$$e_j \approx \frac{\widetilde{\tau_{u,mp}}}{c_m} \left(i\omega c_p - \frac{\kappa_p}{l_j^2}\right) \tag{S128}$$

$$d_j \approx \frac{\left\{\frac{\widetilde{\tau_{u,mp}}}{c_m} \frac{\kappa_p \kappa_m}{l_{j,YIG}^4} - \left[\kappa_p + \kappa_m + i\omega\widetilde{\tau_{u,mp}}\left(\kappa_p + \frac{\kappa_m c_p}{c_m}\right)\right]\frac{1}{l_j^2} + i\omega(c + i\omega\widetilde{\tau_{u,mp}} c_p)\right\}}{\left(\frac{\Gamma_m \sigma_m}{l_j^2} - i\omega \frac{\partial u}{\partial \mu_m}\right)} \tag{S129}$$

and

$$\frac{1}{l_j^6} + \frac{f_2}{l_j^4} + \frac{f_1}{l_j^2} + f_0 = 0 \tag{S130}$$

with

$$f_2 = -\frac{(\kappa_p + \kappa_M) c_m}{\kappa_p \kappa_M \widetilde{\tau_{u,mp}}} - \frac{\kappa_m}{\kappa_M \sigma_m \widetilde{\tau_{n,\mu}}} \frac{\partial n_m}{\partial \mu_m} + i\omega \left[\frac{1}{\kappa_M}\left(S_m \frac{\partial u}{\partial \mu_m} + \Gamma_{m,\mu} \frac{\partial n_m}{\partial T_m}\right) - \frac{\kappa_m}{\kappa_M \sigma_m} \frac{\partial n_m}{\partial \mu_m} - \frac{c_p}{\kappa_p} - \frac{c_m}{\kappa_M}\right] \tag{S131}$$

$$f_1 = \frac{c_m}{\kappa_p \kappa_M \sigma_m \widetilde{\tau_{u,mp}}} \left\{\frac{\partial n_m}{\partial \mu_m} \left(\frac{1}{\widetilde{\tau_{n,\mu}}} + i\omega\right)\left[\kappa_p + \kappa_m + i\omega\widetilde{\tau_{u,mp}}\left(\kappa_p + \frac{\kappa_m c_p}{c_m}\right)\right]\right.$$

$$+ i\omega \left[\sigma_m(c + i\omega\widetilde{\tau_{u,mp}} c_p) - \sigma_m\left(1 + i\omega\widetilde{\tau_{u,mp}} \frac{c_p}{c_m}\right)\left(S_m \frac{\partial u}{\partial \mu_m} + \Gamma_m \frac{\partial n_m}{\partial T_m}\right)\right.$$

$$\left.\left. - i\omega\widetilde{\tau_{u,mp}} \frac{\kappa_p}{c_m} \frac{\partial u}{\partial \mu_m} \frac{\partial n_m}{\partial T_m}\right]\right\} \tag{S132}$$

$$f_0 = \frac{i\omega c_m}{\kappa_p \kappa_M \sigma_m \widetilde{\tau_{u,mp}}} \left[i\omega \frac{\partial n}{\partial T_m} \frac{\partial u}{\partial \mu_m} \left(1 + i\omega\widetilde{\tau_{u,mp}} \frac{c_p}{c_m}\right) - (c + i\omega\widetilde{\tau_{u,mp}} c_p) \frac{\partial n_m}{\partial \mu_m} \left(\frac{1}{\widetilde{\tau_{n,\mu}}} + i\omega\right)\right] \tag{S133}$$

The solution is

$$\frac{1}{l_j} = \left\{ -\frac{1}{3f_3}\left[ f_2 + \left(\frac{-1+i\sqrt{3}}{2}\right)^j C_r + \frac{f_2{}^2 - 3f_3 f_1}{\left(\frac{-1+i\sqrt{3}}{2}\right)^j C_r} \right] \right\}^{\frac{1}{2}} \qquad (S134)$$

where $C_r$ is any one of the three cubic roots that differ from each other by a factor of $e^{i2\pi/3} = \frac{-1+i\sqrt{3}}{2}$ or $e^{i4\pi/3}$

$$C_r = \left\{ \frac{2f_2{}^3 - 9f_1 f_2 f_3 + 27 f_3{}^2 f_0 + \left[\left(2f_2{}^3 - 9f_1 f_2 f_3 + 27 f_3{}^2 f_0\right)^2 - 4\left(f_2{}^2 - 3f_3 f_1\right)^3\right]^{\frac{1}{2}}}{2} \right\}^{\frac{1}{3}} \qquad (S135)$$

The $l_j$ values are controlled by the YIG thermal penetration depth, $\Lambda \equiv \left(\frac{\alpha}{\omega}\right)^{1/2}$, magnon spin diffusion length, $\lambda \equiv \left(\frac{\sigma_m \widetilde{\tau_{n,\mu}}}{\frac{\partial n_m}{\partial \mu_m}}\right)^{1/2}$, and magnon-phonon thermalization length, $l_{mp} \equiv \left[\left(\frac{1}{\kappa_M} + \frac{1}{\kappa_p}\right)\frac{c_m}{\widetilde{\tau_{u,mp}}}\right]^{-1/2}$.

### N. Temperature and Chemical Potential Distributions

The Pt surface ($z = l$) is adiabatic,

$$\hat{Q}_E - \kappa_{e,Pt}\left(\frac{\partial \hat{\vartheta}_{e,Pt}}{\partial z}\right)_{z=l} = 0 \qquad (S136)$$

and

$$\left(\frac{\partial \hat{\vartheta}_{p,Pt}}{\partial z}\right)_{z=l} = 0 \qquad (S137)$$

where the Ettingshausen heat current density $\hat{Q}_E = \kappa_{Pt} \Xi B_z \hat{J}_x$ includes both the ordinary and anomalous contributions, and the Ettingshausen coefficient ($\Xi$) is related to the Nernst ($N$) coefficient according to the Bridgeman relation, $\kappa_{Pt} \Xi = NT$.

The spin flux vanishes at the YIG/GGG interface ($z = -d$),

$$\hat{J}_n = -\sigma_m\left(\nabla \hat{\mu}_m + S_m \nabla \hat{\vartheta}_m\right) = 0 \qquad (S138)$$

The spin mediated heat flux ($Q_{m,GGG}$) and phonon mediated heat flux ($Q_{p,GGG}$) respectively, across the YIG/GGG interface ($z = -d$) are

$$\hat{Q}_{m,GGG} = \Gamma_{m,\mu}\hat{J}_n - \kappa_M \nabla \hat{\vartheta}_m = 0 \tag{S139}$$

and

$$\hat{Q}_{p,GGG} = -\kappa_{p,YIG}\left(\frac{\partial \hat{\vartheta}_{p,YIG}}{\partial z}\right)_{z=-d} = -G_{GGG}\hat{\vartheta}_{p,YIG} \tag{S140}$$

At the Pt/YIG interface ($z = 0$), the spin current continuity is expressed as

$$\hat{J}_n = -\sigma_m(\nabla\hat{\mu}_m + S_m\nabla\hat{\vartheta}_m) = m_2 q_{32} = -\frac{m_2 \hat{J}_s^i}{e} = -g_s\left(\frac{\hat{\mu}_s^i - \hat{\mu}_m^i}{e^2} + \frac{S_s}{e}\hat{\vartheta}_{m-e}^i\right) \tag{S141}$$

At the Pt/YIG interface, the spin mediated heat current density ($Q_s^i$) and the phonon mediated heat current density ($Q_p^i$) are

$$\hat{Q}_s^i = \hat{Q}_E - \kappa_{e,Pt}\left(\frac{\partial \hat{\vartheta}_{e,Pt}}{\partial z}\right)_{z=0} = \hat{J}_{u,m} = \Gamma_{m,\mu}\hat{J}_n - \kappa_M\nabla\hat{\vartheta}_m = \frac{1}{e}g_s S_s T(\hat{\mu}_s^i - \hat{\mu}_m^i) + G_s\hat{\vartheta}_{m-e}^i \tag{S142}$$

and

$$\hat{Q}_p^i = -\kappa_{p,Pt}\left(\frac{\partial \hat{\vartheta}_{p,Pt}}{\partial z}\right)_{z=0} = -\kappa_{p,YIG}\left(\frac{\partial \hat{\vartheta}_{p,YIG}}{\partial z}\right)_{z=0} = G_p(\hat{\vartheta}_{p,YIG,\omega\to 0} - \hat{\vartheta}_{p,Pt,\omega\to 0}) \tag{S143}$$

where $G_s \equiv \left(\frac{\partial \hat{Q}_s^i}{\partial \hat{\vartheta}_{m-e}^i}\right)_{\hat{\mu}_s^i = \hat{\mu}_m^i}$ is close-circuit magnon interface thermal conductance while the open-circuit magnon interface thermal conductance $G_s'$ is defined as $G_s' \equiv \frac{\partial \hat{Q}_{s,\hat{J}_n=0}^i}{\partial \hat{\vartheta}_{m-e,\hat{J}_n=0}^i} = G_s - g_s S_s^2 T \geq 0$. $G_p$ is the phonon contribution to the interface thermal conductance at the Pt/YIG interface. The ten boundary condition equations can be arranged into a matrix form as $WA = S$, where both $S = S^J \hat{q} l + S^I \frac{\Theta_{SH}\hat{I}}{wl} + S^E \hat{Q}_E$ and $A = A^J \hat{q} l + A^I \frac{\Theta_{SH}\hat{I}}{wl} + A^E \hat{Q}_E = W^{-1}S^J \hat{q} l + W^{-1}S^I \frac{\Theta_{SH}\hat{I}}{Wl} + W^{-1}S^E \hat{Q}_E$ contain a single column of ten elements and $W$ is a ten by ten matrix. The elements in $A$ are the ten coefficients in the electron-phonon-magnon temperature solutions according to

$$A_j = a_{j,Pt} \text{ for } j = 0,1 \tag{S144}$$
$$A_j = b_{j-2,Pt} \text{ for } j = 2,3 \tag{S145}$$
$$A_j = a_{j-4} \text{ for } j = 4,5,6 \tag{S146}$$
$$A_j = b_{j-7} \text{ for } j = 7,8,9 \tag{S147}$$

All the elements in $W$ and $S$ can be identified from the ten boundary equations to solve for $A$ via matrix inversion.

The magnon-electron temperature different at the Pt/YIG interface is

$$\hat{\vartheta}^i_{m-e} \equiv \hat{\vartheta}^i_m - \hat{\vartheta}^i_e$$

$$= \sum_{j=0,1,2} (e_j + 1)\left\{a_j \exp\left(\frac{z}{l_j}\right) + b_j \exp\left(\frac{-z}{l_j}\right)\right\} - \sum_{j=0,1} \zeta_{j,Pt} b_{j,Pt}$$

$$+ \left[1 + \frac{i\alpha_{Pt}\kappa_{Pt}}{\alpha_{p,Pt}\kappa_{e,Pt}}\left(\frac{l_{ep}}{\Lambda_{Pt}}\right)^2\right] \frac{i\Lambda_{Pt}^2 \hat{q} l}{\kappa_{Pt} l \left[1 + \frac{i\alpha_{Pt}^2}{\alpha_{p,Pt}\alpha_{e,Pt}}\left(\frac{l_{ep}}{\Lambda_{Pt}}\right)^2\right]}$$

$$= \hat{\vartheta}^{i\;I}_{m-e} \frac{\Theta_{SH}\hat{I}}{Wl} + \hat{\vartheta}^{i\;q}_{m-e} \hat{q} l + \hat{\vartheta}^{i\;E}_{m-e} \hat{Q}_E \tag{S148}$$

where

$$\hat{\vartheta}^{i\;I}_{m-e} = \sum_{j=0,1,2} (e_j + 1)\left\{a_j^I \exp\left(\frac{z}{l_j}\right) + b_j^I \exp\left(\frac{-z}{l_j}\right)\right\} - \sum_{j=0,1} \zeta_{j,Pt} b_{j,Pt}^I \tag{S149}$$

$$\hat{\vartheta}^{i\;q}_{m-e} = \sum_{j=0,1,2} (e_j + 1)\left\{a_j^q \exp\left(\frac{z}{l_j}\right) + b_j^q \exp\left(\frac{-z}{l_j}\right)\right\} - \sum_{j=0,1} \zeta_{j,Pt} b_{j,Pt}^q$$

$$+ \left[1 + \frac{i\alpha_{Pt}\kappa_{Pt}}{\alpha_{p,Pt}\kappa_{e,Pt}}\left(\frac{l_{ep}}{\Lambda_{Pt}}\right)^2\right] \frac{i\Lambda_{Pt}^2}{\kappa_{Pt} l \left[1 + \frac{i\alpha_{Pt}^2}{\alpha_{p,Pt}\alpha_{e,Pt}}\left(\frac{l_{ep}}{\Lambda_{Pt}}\right)^2\right]} \tag{S150}$$

$$\hat{\vartheta}^{i\;E}_{m-e} = \sum_{j=0,1,2} (e_j + 1)\left\{a_j^E \exp\left(\frac{z}{l_j}\right) + b_j^E \exp\left(\frac{-z}{l_j}\right)\right\} - \sum_{j=0,1} \zeta_{j,Pt} b_{j,Pt}^E \tag{S151}$$

The magnon chemical potential at Pt/YIG interface is

$$\hat{\mu}^i_m = \sum_{j=0,1,2} d_j \left\{a_j \exp\left(\frac{z}{l_j}\right) + b_j \exp\left(\frac{-z}{l_j}\right)\right\} = \hat{\mu}^{i\;I}_m \frac{\Theta_{SH}\hat{I}}{Wl} + \hat{\mu}^{i\;q}_m \hat{q} l + \hat{\mu}^{i\;E}_m \hat{Q}_E \tag{S152}$$

where

$$\hat{\mu}^{i\;I}_m = \sum_{j=0,1,2} d_j \left\{a_j^I \exp\left(\frac{z}{l_j}\right) + b_j^I \exp\left(\frac{-z}{l_j}\right)\right\} \tag{S153}$$

$$\hat{\mu}^{i\;q}_m = \sum_{j=0,1,2} d_j \left\{a_j^q \exp\left(\frac{z}{l_j}\right) + b_j^q \exp\left(\frac{-z}{l_j}\right)\right\} \tag{S154}$$

$$\hat{\mu}_m^{i\ E} = \sum_{j=0,1,2} d_j \left\{ a_j^E \exp\left(\frac{z}{l_j}\right) + b_j^E \exp\left(\frac{-z}{l_j}\right) \right\} \quad (S155)$$

The Pt phonon temperature modulation is similarly separated into different components as

$$\hat{\vartheta}_{p,Pt} = \sum_{j=0,1} \left[ a_{j,Pt} \sinh\left(\frac{z}{l_{j,Pt}}\right) + b_{j,Pt} \cosh\left(\frac{z}{l_{j,Pt}}\right) \right] - \frac{i\Lambda_{Pt}^2 \hat{q}_\omega l}{\kappa_{Pt} l \left[ 1 + \frac{i\alpha_{Pt}^2}{\alpha_{p,Pt}\alpha_{e,Pt}} \left(\frac{l_{ep}}{\Lambda_{Pt}}\right)^2 \right]}$$

$$= \hat{\vartheta}_{p,Pt}^{I} \frac{\Theta_{SH}\hat{I}}{Wl} + \hat{\vartheta}_{p,Pt}^{q} \hat{q}l + \hat{\vartheta}_{p,Pt}^{E} \hat{Q}_E \quad (S156)$$

where

$$\hat{\vartheta}_{p,Pt}^{I} = \sum_{j=0,1} \left[ a_{j,Pt}^{I} \sinh\left(\frac{z}{l_{j,Pt}}\right) + b_{j,Pt}^{I} \cosh\left(\frac{z}{l_{j,Pt}}\right) \right] \quad (S157)$$

$$\hat{\vartheta}_{p,Pt}^{q} = \sum_{j=0,1} \left[ a_{j,Pt}^{q} \sinh\left(\frac{z}{l_{j,Pt}}\right) + b_{j,Pt}^{q} \cosh\left(\frac{z}{l_{j,Pt}}\right) \right] - \frac{i\Lambda_{Pt}^2}{\kappa_{Pt} l \left[ 1 + \frac{i\alpha_{Pt}^2}{\alpha_{p,Pt}\alpha_{e,Pt}} \left(\frac{l_{ep}}{\Lambda_{Pt}}\right)^2 \right]} \quad (S158)$$

$$\hat{\vartheta}_{p,Pt}^{E} = \sum_{j=0,1} \left[ a_{j,Pt}^{E} \sinh\left(\frac{z}{l_{j,Pt}}\right) + b_{j,Pt}^{E} \cosh\left(\frac{z}{l_{j,Pt}}\right) \right] \quad (S159)$$

Unitary Fourier transform is used to obtain $\hat{I}$ and $\hat{q}$. Inverse Unitary Fourier transform, $f(t) = \frac{1}{\sqrt{2\pi}} \int_{-\infty}^{\infty} \hat{f}(\omega) e^{-i\omega t} d\omega$, is then used to obtain the following time-domain expressions

$$f(t) = \langle \hat{f}_{\omega \to 0} \rangle^q \frac{I_{\omega_J,X}^2}{\sigma_{xx} W^2 l}$$

$$+ \sqrt{2} \left[ f_{\omega_J,X} \sin\omega_J t + f_{\omega_J,Y} \cos\omega_J t + f_{2\omega_J,X} \sin 2\omega_J t + f_{2\omega_J,Y} \cos 2\omega_J t \right] \quad (S160)$$

where $f$ is either $\langle \vartheta_{p,Pt} \rangle$, $\vartheta_{m-e}^i$, or $\hat{\mu}_m^i$.

Therefore,

$$\langle \vartheta_{p,Pt,\omega_J,X} \rangle = \left[ \text{Re}\left( \langle \hat{\vartheta}_{p,Pt,\omega_J}^{I} \rangle \right) \Theta_{SH} + \text{Re}\left( \langle \hat{\vartheta}_{p,Pt,\omega_J}^{E} \rangle \right) \kappa_{Pt} \Xi B_z \right] \frac{I_{\omega_J,X}}{Wl} \quad (S161)$$

$$\langle \vartheta_{p,Pt,\omega_J,Y} \rangle = \left[ \text{Im}\left( \langle \hat{\vartheta}_{p,Pt,\omega_J}^{I} \rangle \right) \Theta_{SH} + \text{Im}\left( \langle \hat{\vartheta}_{p,Pt,\omega_J}^{E} \rangle \right) \kappa_{Pt} \Xi B_z \right] \frac{I_{\omega_J,X}}{Wl} \quad (S162)$$

$$\langle \vartheta_{p,Pt,2\omega_J,X} \rangle = \text{Im}\left[ \langle \hat{\vartheta}_{p,Pt,2\omega_J}^{q} \rangle \right] \frac{I_{\omega_J,X}^2}{\sqrt{2}\sigma_{xx} W^2 l} \quad (S163)$$

$$\langle \vartheta_{p,Pt,2\omega_J,Y}\rangle = -\text{Re}\left[\langle \hat{\vartheta}_{p,Pt,2\omega_J}{}^q\rangle\right]\frac{I_{\omega_J,X}{}^2}{\sqrt{2}\sigma_{xx}W^2 l} \tag{S164}$$

$$\vartheta_{m-e,2\omega_J,X}^i = \text{Im}\left[\hat{\vartheta}_{m-e,2\omega_J}^i{}^q\right]\frac{I_{\omega_J,X}{}^2}{\sqrt{2}\sigma_{xx}W^2 l} \tag{S165}$$

$$\vartheta_{m-e,2\omega_J,Y}^i = -\text{Re}\left[\hat{\vartheta}_{m-e,2\omega_J}^i{}^q\right]\frac{I_{\omega_J,X}{}^2}{\sqrt{2}\sigma_{xx}W^2 l} \tag{S166}$$

$$\mu_{m,2\omega_J,X}^i = \text{Im}\left[\hat{\mu}_{m,2\omega_J}^i{}^q\right]\frac{I_{\omega_J,X}{}^2}{\sqrt{2}\sigma_{xx}W^2 l} \tag{S167}$$

$$\mu_{m,2\omega_J,Y}^i = -\text{Re}\left[\hat{\mu}_{m,2\omega_J}^i{}^q\right]\frac{I_{\omega_J,X}{}^2}{\sqrt{2}\sigma_{xx}W^2 l} \tag{S168}$$

Because $\vartheta_{m-e,2\omega_J}^i$ and $\mu_{m,2\omega_J}^i$ are insensitive to the $\mathbf{H}$ direction,

$$\Delta V_{SSE,2\omega_J}^T(\varphi = 180°) \equiv V_{SSE,2\omega_J}^T(|\mathbf{H}|,\varphi = 180°) - V_{SSE,2\omega_J}^T(-|\mathbf{H}|,\varphi = 180°)$$

$$= -2w_H \frac{g_s\eta\Theta_{SH}\lambda_{Pt}}{\sigma_{11}l}\tanh\left(\frac{l}{2\lambda_{Pt}}\right)m_2\left(S_s\vartheta_{m-e,2\omega_J}^i - \frac{1}{e}\mu_{m,2\omega_J}^i\right) \tag{S169}$$

$$\Delta V_{SSE,2\omega_J,X}^T(\varphi = 180°) = -\sqrt{2}w_H \frac{g_s\eta\Theta_{SH}\lambda_{Pt}I_{\omega_J,X}{}^2}{\sigma_{xx}{}^2 w^2 l^2}\tanh\left(\frac{l}{2\lambda_{Pt}}\right)m_2\text{Im}\left[S_s\hat{\vartheta}_{m-e}^i{}^q - \frac{1}{e}\hat{\mu}_m^i{}^q\right] \tag{S170}$$

$$\Delta V_{SSE,2\omega_J,Y}^T(\varphi = 180°) = \sqrt{2}w_H \frac{g_s\eta\Theta_{SH}\lambda_{Pt}I_{\omega_J,X}{}^2}{\sigma_{xx}{}^2 w^2 l^2}\tanh\left(\frac{l}{2\lambda_{Pt}}\right)m_2\text{Re}\left[S_s\hat{\vartheta}_{m-e}^i{}^q - \frac{1}{e}\hat{\mu}_m^i{}^q\right] \tag{S171}$$

## O. Thermal Conductance of the GGG Substrate

The thermal penetration depth due to Joule heating is reduced to the YIG thickness ($d$) as $\omega$ is increased to $\omega_c = \alpha_{YIG}/d^2 = 24392$ Hz. Therefore, the measured second-harmonic modulation of the Pt phonon temperature, $\langle \vartheta_{p,Pt,2\omega_J}\rangle_{meas}$, is controlled by $\kappa_{YIG}$ and $c_{YIG}$ at $\omega > \omega_c$ and becomes sensitive to $G_{GGG}$ at the low-frequency range with $\omega < \omega_c/2$. These expected behaviors are used to extract ($\kappa_{YIG}, c_{YIG}$) and $G_{GGG}(\omega < \omega_c/2)$ by minimizing the following fitting error between the theoretical $\langle \vartheta_{p,Pt,2\omega_J}\rangle_t$ and the measured $\langle \vartheta_{p,Pt,2\omega_J}\rangle_{meas}$ at $2\omega_J$ above $\omega_c$ and below $\omega_c/2$, respectively,

$$\epsilon_{2\omega_J<\omega_c/2} \equiv \sqrt{\frac{\left(\langle\vartheta_{p,Pt,2\omega_J,Y}\rangle_{meas} - \langle\vartheta_{p,Pt,2\omega_J,Y}\rangle_t\right)^2 + \left(\langle\vartheta_{p,Pt,2\omega_J,X}\rangle_{meas} - \langle\vartheta_{p,Pt,2\omega_J,X}\rangle_t\right)^2}{\langle\vartheta_{p,Pt,2\omega_J,Y}\rangle^2_{meas} + \langle\vartheta_{p,Pt,2\omega_J,X}\rangle^2_{meas}}} \quad (S172)$$

Figure S8 shows the fitting error as a function of the real and imaginary components of $G_{GGG}(\omega)$ at four representative frequencies. Figure S9 shows the extracted $G_{GGG}(\omega < \omega_c/2)$ values. Polynomial fitting of the extracted $G_{GGG}(\omega < \omega_c/2)$ is used to obtain $G_{GGG}(\omega > \omega_c)$. Figure S10 compares the measured $\langle\vartheta_{p,Pt,2\omega_J}\rangle$ and the theoretical results calculated with the as-obtained $G_{GGG}(\omega)$ for the entire frequency range.

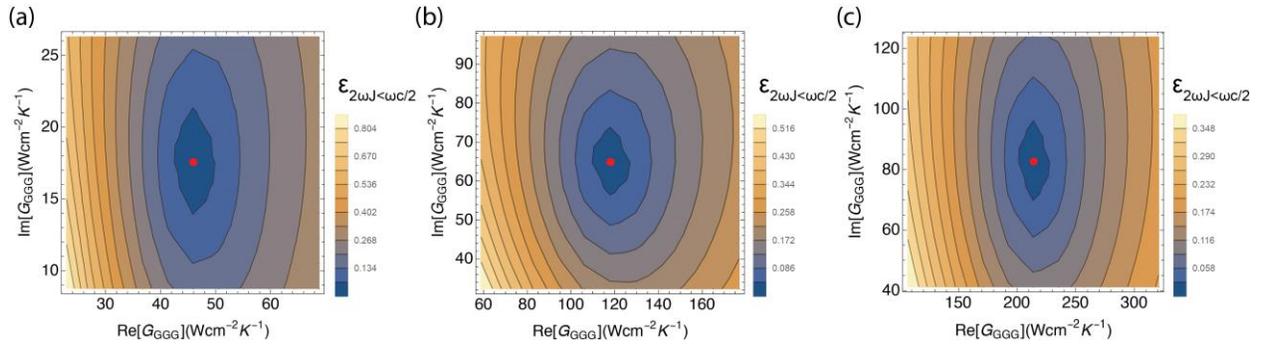

**FIG. S8**. Root-mean-square fitting error as a function of the real and imaginary component of $G_{GGG}(2\omega_J)$ for $\nu_J$ of (a) 150 Hz, (b) 3168 Hz, and (c) 9776 Hz. The thermal penetration depth of YIG at $2\omega_J$ is 39.4 μm, 8.6 μm, and 4.9 μm, for (a), (b), and (c) respectively. The red dot in each contour map identifies the $G_{GGG}(2\omega_J)$ value that minimizes the fitting error.

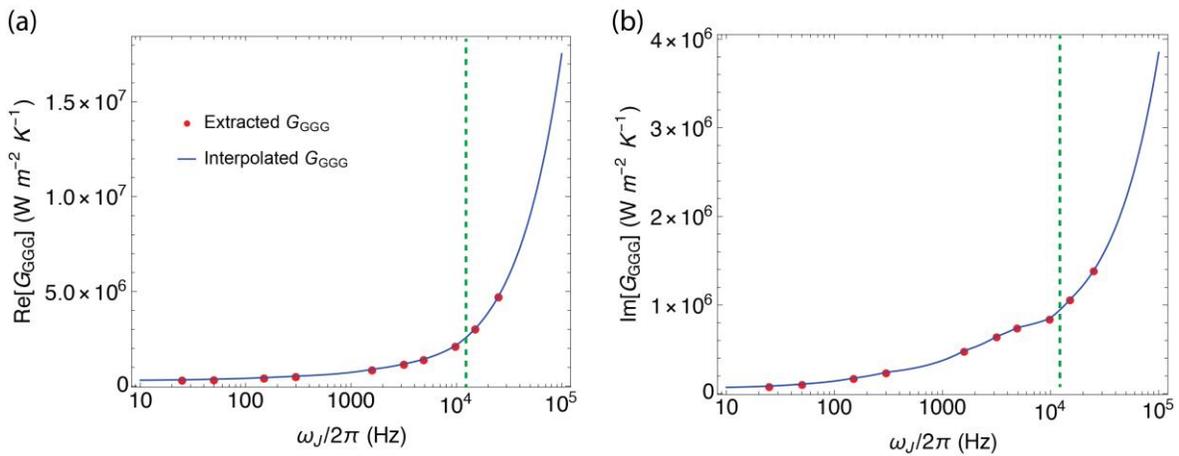

**FIG. S9**. Extracted $G_{GGG}(2\omega_J < \omega_c/2)$ (symbols) at each frequency $\omega_J/2\pi$. The solid line is polynomial fitting of the extracted data. The vertical green dashed line intercepts $\frac{\omega_J}{2\pi} = \nu_J = 12196$ Hz that yields $2\omega_J = \omega_c/2$.

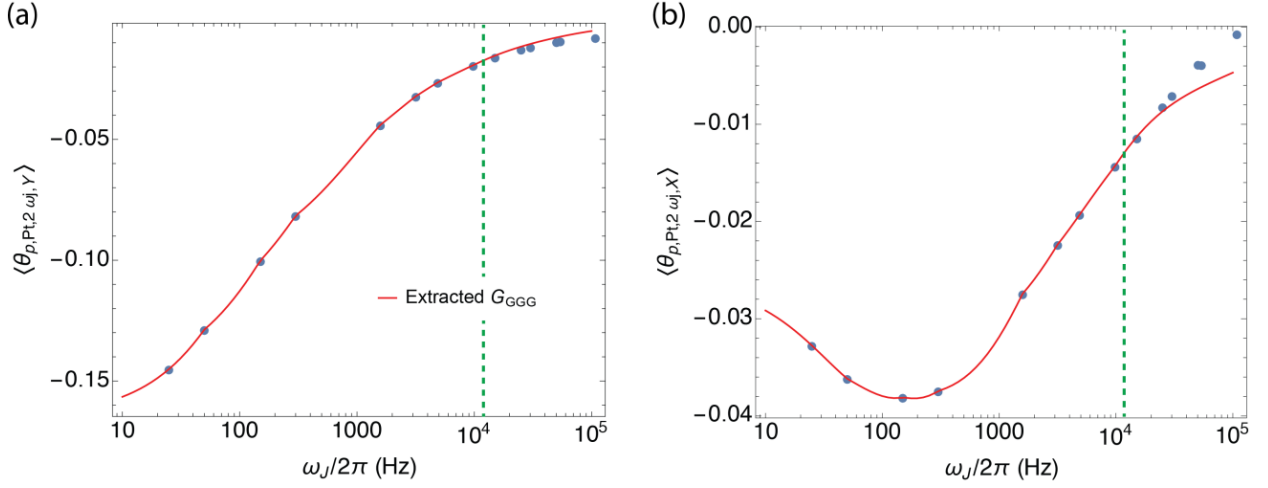

**FIG. S10**. Comparison between the analytical model with the measurement results of the second harmonic modulation of the Pt phonon temperature. (a) The in-phase (Y) and (b) out-of-phase (X) components of the measured (symbols) and analytical results (lines) as a function of the current frequency in the Pt. The vertical green dashed line intercepts $\frac{\omega_J}{2\pi} = \nu_J = 12196$ Hz that yields $2\omega_J = \omega_c/2$.

**P. Comparison between the Theoretical Models with the Measurement Results**

We verify that the analytical model produce several expected behaviors including $\frac{\partial \hat{Q}^i_{p,2\omega_J \to 0}}{\partial (\hat{q}l)} < 0$, $\frac{\partial \hat{Q}^i_{s,2\omega_J \to 0}}{\partial (\hat{q}l)} < 0$, $\frac{\partial \hat{J}_{n,2\omega_J \to 0}}{\partial (\hat{q}l)} < 0$, $\frac{\partial \hat{Q}^i_{s,\omega_J \to 0}}{\partial \left(m_2 \Theta_{SH} \frac{I_{\omega_J,X}}{Wl}\right)} = -\frac{\partial \hat{Q}^i_{p,\omega_J \to 0}}{\partial \left(m_2 \Theta_{SH} \frac{I_{\omega_J,X}}{Wl}\right)} < 0$, and $\frac{\partial \hat{J}_{n,\omega_J \to 0}}{\partial \left(m_2 \Theta_{SH} \frac{I_{\omega_J,X}}{Wl}\right)} > 0$. In addition, we expect $\hat{\vartheta}^{i\ q}_{m-e} < 0$, $\frac{\hat{\vartheta}^{i\ I}_{m-e}}{m_2} < 0$, $\frac{(\mu_s^I - \mu_m^I)^i}{m_2} < 0$, and $(\mu_s^q - \mu_m^q)^i > 0$, where the $i$ superscript is used to represent the value at the Pt/YIG interface. We find a range of magnon transport properties that satisfy these constraints and match the low-frequency amplitudes of the SSE measurement results at 50 Hz as well as the following measured normalized frequency dependence at a frequency near $10^5$ Hz,

$$\Delta V^T_{SSE,2\omega_J}{}^*(\varphi = 180°) = \frac{\Delta V^T_{SSE,2\omega_J}(\varphi = 180°)}{\Delta V^T_{SSE,2\omega_J \to 0,Y}(\varphi = 180°)}. \quad (S173)$$

Table S1 lists the materials and interface properties involved in the analytical model together with a comparison between the literature values and those that are either measured in this work or used to fit the measurement results. Standard methods have yet to be established for measuring the magnon-mediated properties, including the $\lambda$, $l_{mp}$, $\sigma_m$, $S_m$, $\kappa_M$, $c_m$, $S_s$, and $G_s'$. The magnon dispersion in YIG have been approximated in the literature [23] to calculate the $c_m$ values listed in Table S1. In addition, we use the literature analytical models [3,16] to calculate

$$G_s' = \frac{h}{e^2} \frac{k_B T}{\hbar} \frac{\mu_B k_B g_r}{\pi M_s \Lambda^3} = 8.08 \times 10^6 \; (\mathrm{Wm^{-2}K^{-1}}) \quad (S174)$$

Based on a quadratic magnon dispersion and the same average relaxation rate for the magnon momentum current and heat currents, a literature model gives $S_m = 42.36 \; \left(\frac{\mu V}{K}\right)$ [3]. Besides the expectation of $\tau_c < \tau_{mp} < \tau_\mu$, we expect that small-angle inelastic magnon scattering processes relax the heat current faster than the particle current, similar to the behavior of inelastic electron scattering that suppresses the Lorenz ratio [14]. Thus, we require

$$\tau_{\kappa,c} \leq \tau_{\sigma,c} < \tau_{u,mp} = l_{mp}{}^2 c_m \left(\frac{1}{\kappa_M} + \frac{1}{\kappa_p}\right) < \tau_{n,\mu} = \frac{\lambda^2}{\sigma_m} \frac{\partial n_m}{\partial \mu_m} \quad (S175)$$

where $\tau_{\kappa,c}$ and $\tau_{\sigma,c}$ are the average $\left(\overline{\tau_c}^{-1} + \overline{\tau_{mp}}^{-1} + \overline{\tau_\mu}^{-1}\right)^{-1}$ values that appear in Equations S105 and S90 of $\kappa_m$ and $\sigma_m$ and their analytical expressions given in the literature [3]. For given $l_{mp}$ and $\lambda$, Equation S175 provides a constraint for the $\kappa_M$ and $\sigma_m$ values calculated from the literature analytical expressions. Satisfying these constraints, the property values listed in Table S1 can fit the measured $\Delta V_{SSE,2\omega_J}^T$ results.

**TABLE S1. List of transport properties that are from literature estimates, the present measurements, or used in this work for fitting the measurement data in Figs. 3 and 4**

|  | Symbol (Unit) | Pt | YIG |
|---|---|---|---|
| Total thermal conductivity | $\kappa$ (W m$^{-1}$K$^{-1}$) | 20.5 (This work) | 6.5 [15], 7.4 [16], 8.0 [17], 8.5 [18], 8.5 (this work) |
| Electronic thermal conductivity | $\kappa_e$ (W m$^{-1}$K$^{-1}$) | 13.5 (This work) | - |
| Open-circuit magnon thermal conductivity | $\kappa_M$ (W m$^{-1}$K$^{-1}$) | - | 0.1 (this work) |

| Close-circuit magnon thermal conductivity | $\kappa_m$ (W m$^{-1}$K$^{-1}$) | - | 2.7 (this work) |
|---|---|---|---|
| Total specific heat | $c$ (J m$^{-3}$K$^{-1}$) | $2.8 \times 10^6$ [19] | $3.2 \times 10^6$ [16] |
| Electronic specific heat | $c_e$ (J m$^{-3}$K$^{-1}$) | $1.2 \times 10^5$ [20] | - |
| Magnon specific heat | $c_m$ (J m$^{-3}$K$^{-1}$) | - | $5.17 \times 10^4$ [21] |
| Pt spin diffusion length | $\lambda_{Pt}$ (nm) | 1.5 [3,5,10] | - |
| Pt spin Hall angle | $\Theta_{SH}$ | 0.11 [4,5,10] | - |
| YIG magnon spin diffusion length | $\lambda$ (μm) | - | 9.4 (SPE), 8.7 (SSE) [22]<br>3.5 [23], 1.3 [24]<br>4.5 (SPE), 9.0 (SSE) [25]<br>0.65 (SPE), 0.83 (SSE) (This work) |
| Electron-phonon thermalization length | $l_{ep}$ (Pt) (nm) | 2.8 | - |
| Magnon-phonon thermalization length | $l_{mp}$ (YIG) (nm) | - | 6 [3,5,10], 300 [12,26]<br>250 (This work) |
| Pt conductivity | $\sigma_{xx}$ ($\Omega^{-1}$m$^{-1}$) | $1.85 \times 10^6$ (This work) | - |
| YIG magnon conductivity | $\sigma_m$ ($\Omega^{-1}$m$^{-1}$) | - | $5 \times 10^5$ [3], $1.1 \times 10^6$ [12]<br>$2.15 \times 10^6$ (This work) |
| Bulk magnon spin Seebeck coefficient | $S_m$ (μV K$^{-1}$) | - | 42.4 [3]<br>55 (SPE), 63.5 (SSE) (This work) |
| Interface spin Seebeck coefficient | $S_s$ (μV K$^{-1}$) | 64.6 [5],<br>168.0 (SPE), 65.5 (SSE) (This work) | |

| Interface phonon thermal conductance | $G_P$ (W m$^{-2}$K$^{-1}$) | 2.0 ~ 2.8 × 10$^8$ [5,10], 1.7 × 10$^8$ (This work) |
|---|---|---|
| Open-circuit Interface magnon thermal conductance | $G_s'$ (W m$^{-2}$K$^{-1}$) | 8.08 × 10$^6$ (Calculated from the model [3,5,10,27]) |
| Real component of the interfacial spin mixing conductance | $g_r$ ($\Omega^{-1}$m$^{-2}$) | - 1.6 × 10$^{14}$ [3]<br>7.0 × 10$^{14}$ [5]<br>1.11 × 10$^{14}$ (This work) |
| Ratio for the effective spin conductance | $\mathfrak{R} \equiv g_s/g_r$ | - 0.06 [3]<br>0.16 [5], (This work) |

When λ is increased to exceed Λ at high current modulation frequency of 100 kHz, the anti-crossing behavior in $l_0$, $l_1$, and $l_2$ (Fig. 4b) is accompanied with the emergence of wavelike profiles in both the temperatures and $\mu_m$ (Fig. S11). As shown in Figure S12, the ratio $U_{sc} \equiv \frac{\Delta V_{2\omega_J \to 0,Y}^T}{\langle \vartheta_{p,Pt,\omega_J \to 0,X} \rangle}$ calculated from the model is insensitive to most of the transport properties except for a weak dependence on the spin diffusion length ($\lambda_{Pt}$) in Pt. This behavior reveals the Onsager reciprocity between the SSE and SPE coefficients at the same magnon chemical potential and temperature. Because the magnon chemical potential and temperature can be different in the SSE and SPE measurements, the SPE coefficient can deviate from that calculated from the Onsager reciprocity. With this effect taken into account, a modest change of two fitting parameters $S_s = 168\ \mu V/K$ and $S_m = 55\ \mu V/K$ would allow the model fitting of $\langle \vartheta_{p,Pt,\omega_J \to 0} \rangle$ (Fig. S13).

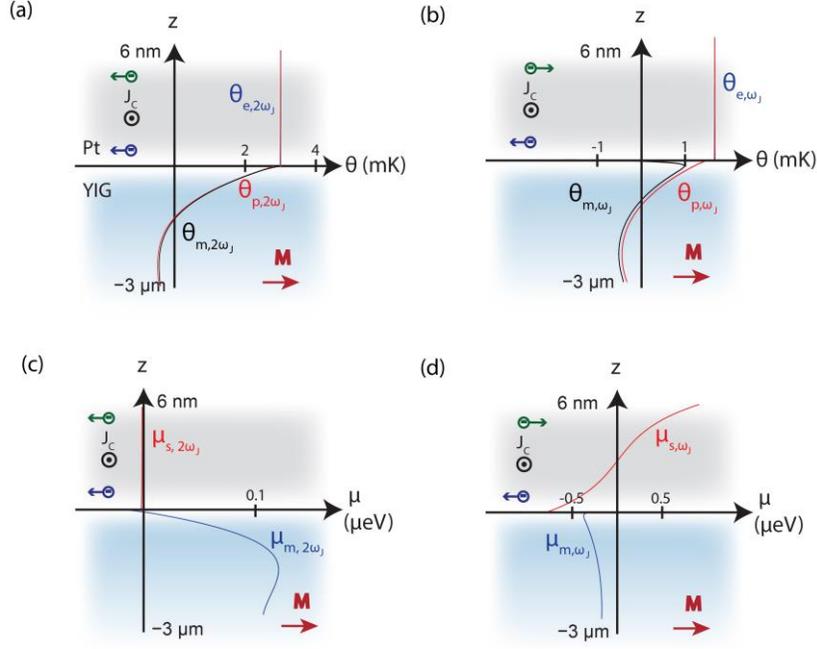

**FIG. S11**. Analytical solutions of second harmonic (a,c) and first harmonic (b,d) temperature and spin accumulation modulations for Pt electrons ($\theta_e$, $\mu_s$), Pt and YIG phonons ($\theta_p$), and YIG magnons ($\theta_m$, $\mu_m$) at a high current modulation frequency of 100 kHz when $\lambda = 2.5$ μm.

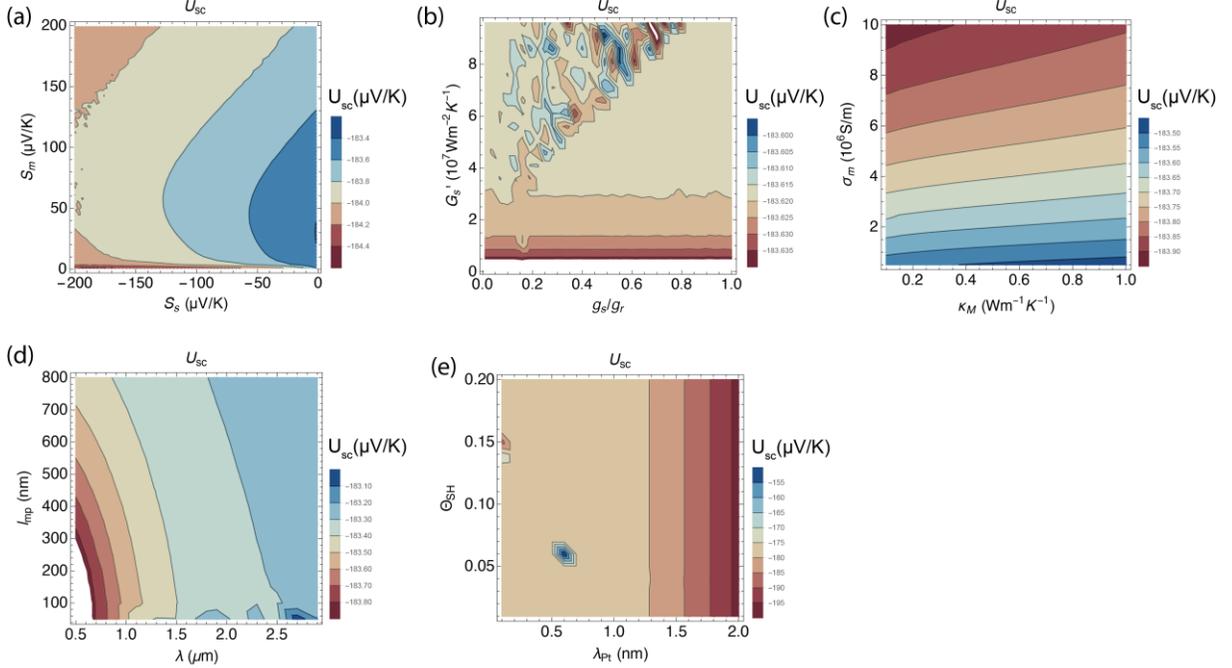

**FIG. S12**. Contour plots of $U_{sc}$ as a function of (a) $S_s - S_m$, (b) $G_s' - \mathfrak{R}$, (c) $\sigma_m - \kappa_M$, (d) $l_{mp} - \lambda$, and (e) $\Theta_{SH} - \lambda_{Pt}$. The interval of each contour is less than 1 % of $U_{sc}$ value except (e).

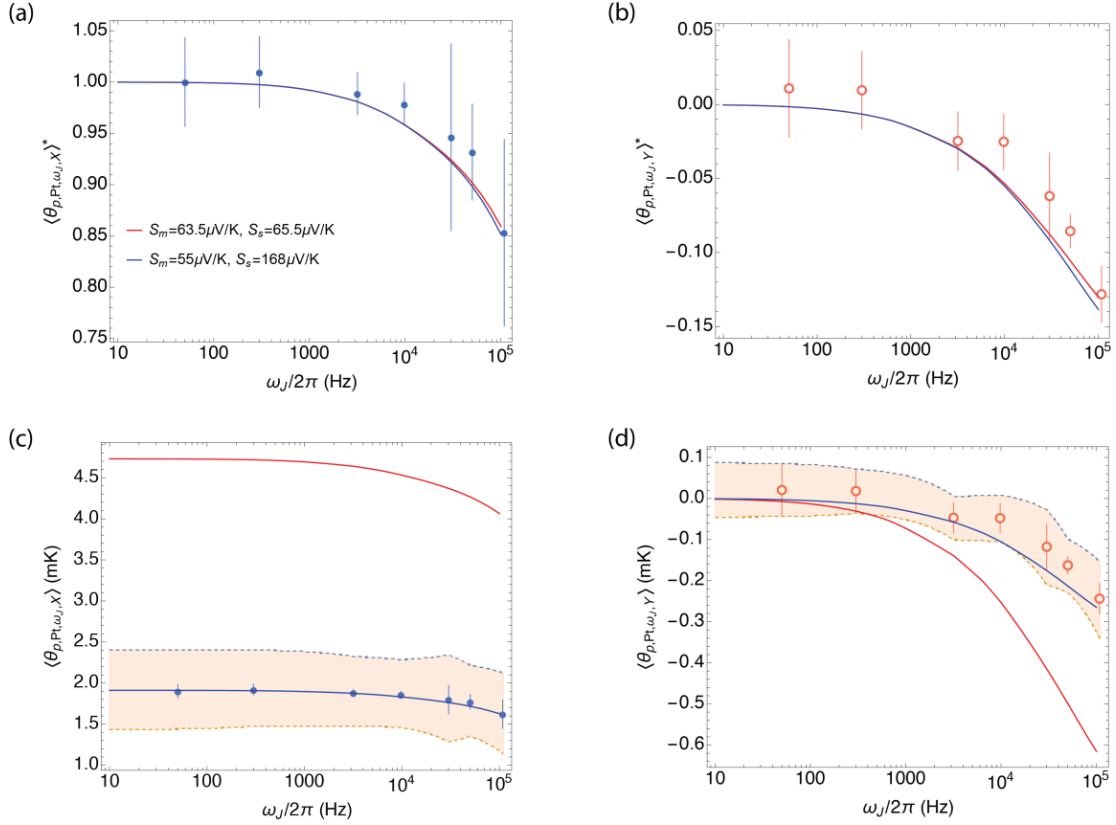

**FIG. S13.** Comparison between experimental data (symbols) and the analytical model (solid lines) of (a,b) $\langle \vartheta_{p,Pt,\omega_J} \rangle^*$ and (c,d) $\langle \vartheta_{p,Pt,\omega_J} \rangle$. The red lines are calculated based on the fitting parameters obtained from $\Delta V_{SSE,2\omega_J}^T$ fitting, while the blue lines are obtained by using $S_s = 168\ \mu V/K$ and $S_m = 55\ \mu V/K$ to fit the $\langle \vartheta_{p,Pt,\omega_J \to 0} \rangle$ magnitude without changing other properties listed in Table 1 for the $\Delta V_{SSE,2\omega_J}^T$ fitting.

The value of $G_P \approx (1.7 \pm 0.5) \times 10^8$ Wm$^{-2}$K$^{-1}$ used in our model calculation is based on the $G_P$ value measured by David Cahill group on a similar sputtered Pt/YIG interface. This $G_P$ value is close to the theoretical values reported in the literature [5,10]. The corresponding interface Kapitza length $\ell_K \equiv \frac{\kappa_{YIG}}{G_P} \approx 60 \pm 20$ nm, which is still much smaller than both the YIG film thickness $d$ and the thermal penetration depth $\Lambda$. Since the $G_P$ value can vary depends on interface quality and growth conditions, we investigated the effect of $G_P$ value on the frequency dependence of SSE and SPE signal (Fig. S14). Varying the $G_P$ value by one order of magnitude only causes a change in the magnitude and negligible change in the frequency dependence for both the SSE and SPE signals.

With the use of the property values listed in Table S1, we investigated the effect of YIG film thickness $d$ on the analytical model results. As shown in Fig. S15, both the frequency dependence and the low-frequency magnitude of both the SSE and SPE signals are reduced with decreasing $d$, consistent with the reported experimental observations [28,29]. The decreased low-frequency magnitude and frequency dependence is due to a reduced $d$ toward the magnon spin diffusion length and thermal penetration depth.

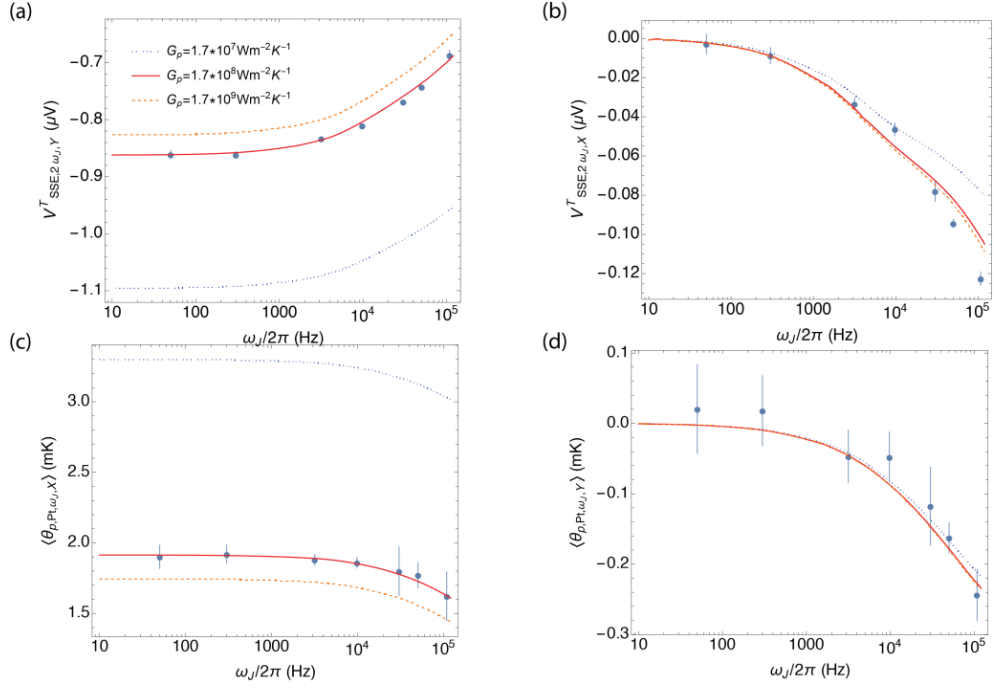

**FIG. S14**. Comparison between the experimental data (symbols) and the analytical model (solid lines) of $\Delta V^T_{SSE,2\omega_J}$ and $\langle \vartheta_{p,Pt,\omega_J} \rangle$ at different $G_p$ values while the values for other parameters are the same as those used for fitting $\Delta V^T_{SSE,2\omega_J}$ and $\langle \vartheta_{p,Pt,\omega_J} \rangle$, respectively.

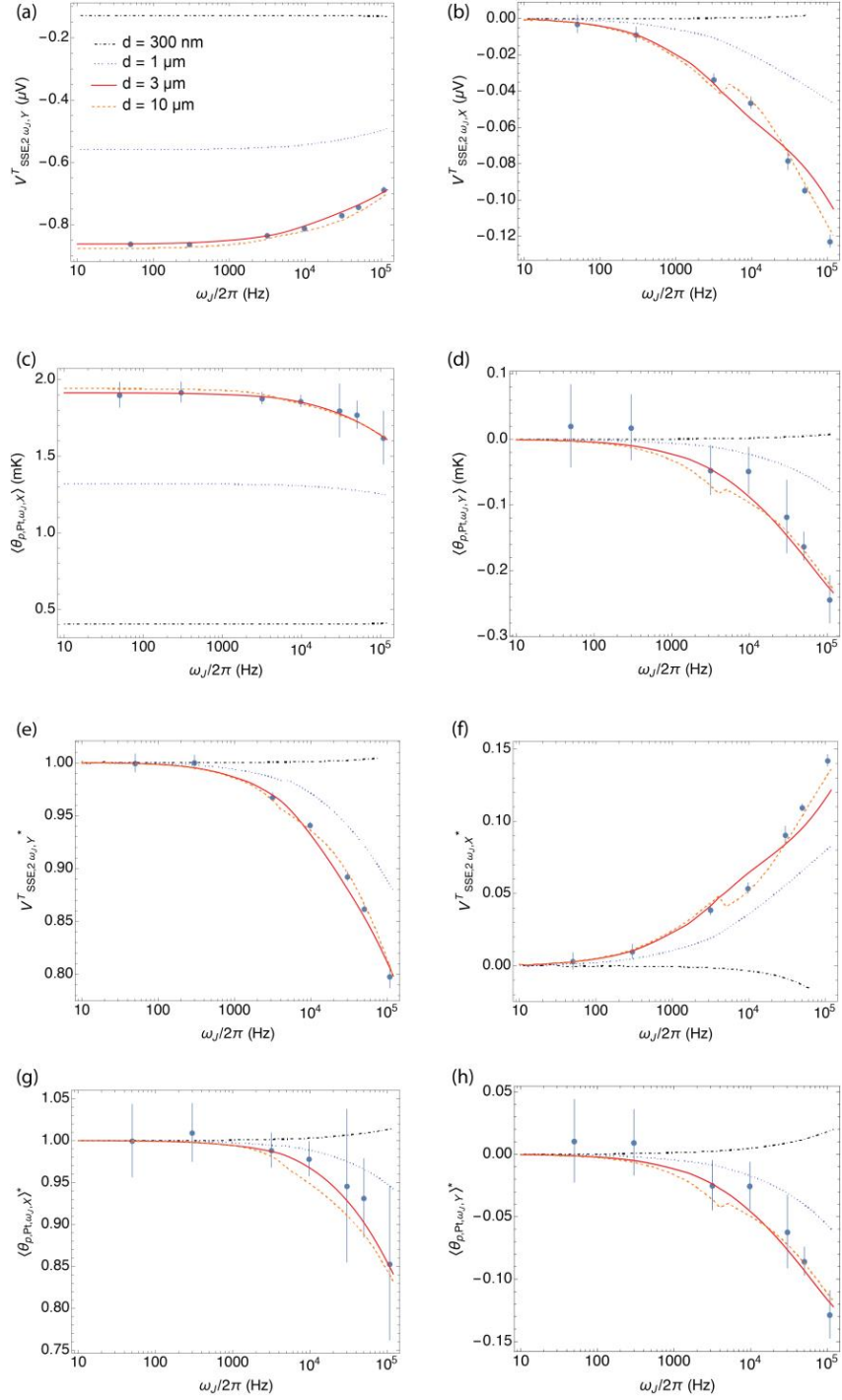

**FIG. S15**. Comparison between the experimental data (symbols) and the analytical model (lines) of dimensional and normalized $\Delta V^T_{SSE,2\omega_J}$ (a, b, e, f) and $\langle \vartheta_{p,Pt,\omega_J} \rangle$ (c, d, g, h) at different YIG film thickness ($d$) values shown in Panel a while the values for other parameters are the same as those used for fitting $\Delta V^T_{SSE,2\omega_J}$ and $\langle \vartheta_{p,Pt,\omega_J} \rangle$, respectively.

## Q. Anomalous Ettingshausen Effect

The Ettingshausen heat current, including the anomalous Ettingshausen effect (ANE) caused by proximity-induced magnetization in Pt, is carried by electrons in the Pt. Without a spin-mediated interface heat transport channel, both the electron heat flux and the phonon heat flux, need to vanish at steady state at the Pt-YIG interface just like at the top Pt surface, so as to satisfy energy conservation in the Pt film. Only when electrons in the Pt and magnons in the YIG can exchange both angular momentum and heat can phonons at the two sides exchange energy at the same rate and opposite direction as the spin-mediated interfacial transfer of electronic heat. As such, our model calculation in Fig. S16 show that ANE is unable to produce the observed frequency dependence in the UMR signal that is caused by the reduced thermal penetration depth in YIG compared to the magnon spin diffusion length and magnon-phonon thermalization length. Meanwhile, an extremely large Nernst coefficient of the Pt, $N = 7.6 \times 10^{-5}$ VK$^{-1}$T$^{-1}$, is required to produce the magnitude of the observed UMR effect. This $N$ value is more than three orders of magnitude larger than the $1.6 \times 10^{-8}$ VK$^{-1}$T$^{-1}$ value for CoFeB. Therefore, the ANE induced in Pt due to its proximity to a FI is unlikely to make a major contribution to the measured $\langle \vartheta_{p,Pt,\omega_J} \rangle$, in agreement with prior report [10].

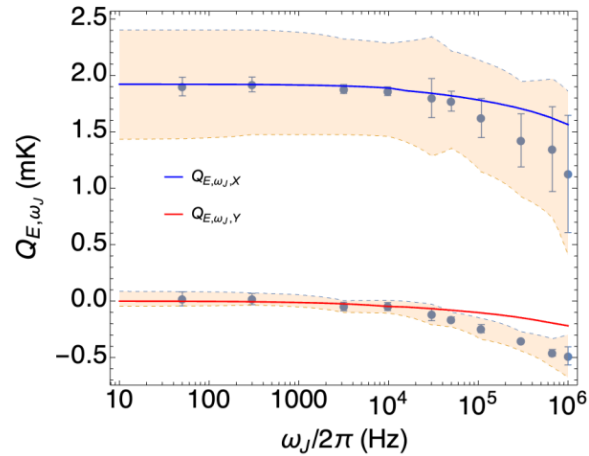

**FIG. S16**. Comparison between experimental $\langle \vartheta_{p,Pt,\omega_J} \rangle$ data (symbols) and the analytical solution (solid lines) of $\hat{Q}_{E,\omega_J}$ when the $N = 7.6 \times 10^{-5}$ VK$^{-1}$T$^{-1}$.